\documentclass[12pt]{article}
\pdfoutput=1
\usepackage{jheppub}
\usepackage{pdflscape,float}

\newlength{\abstractwidth}
\usepackage{amsthm}
\usepackage{amsmath}
\usepackage{amsfonts}
\usepackage{amssymb}
\usepackage{latexsym}
\usepackage{wasysym}

\usepackage{subcaption,siunitx,booktabs}
\usepackage{epsf}
\usepackage{color}
\usepackage{graphicx}
\usepackage{tikz}
\usepackage{dsfont}

\usetikzlibrary{arrows,shapes,positioning}
\usetikzlibrary{decorations.markings}
\usepackage[rightcaption]{sidecap}
\tikzstyle arrowstyle=[scale=1]
\tikzstyle directed=[postaction={decorate,decoration={markings,
    mark=at position .65 with {\arrow[arrowstyle]{stealth}}}}]
\tikzstyle reverse directed=[postaction={decorate,decoration={markings,
    mark=at position .65 with {\arrowreversed[arrowstyle]{stealth};}}}]
\usetikzlibrary{positioning}

\tikzset{line/.style={line width=0.25mm},
curve/.style={line,smooth,tension=1},
->-/.style={decoration={
  markings,
  mark=at position #1 with {\arrow[>=stealth]{>}}},postaction={decorate}},
-<-/.style={decoration={
  markings,
  mark=at position #1 with {\arrow[>=stealth]{<}}},postaction={decorate}},
}

\usepackage{hyperref}
\definecolor{darkred}{rgb}{0.8,0.1,0.1}
\hypersetup{colorlinks=true, linkcolor=darkred, citecolor=blue, linktoc=page}

\renewcommand{\thanks}[1]{\footnote{#1}}

\newcommand{\bea}{\begin{eqnarray}}
\newcommand{\eea}{\end{eqnarray}}
\newcommand{\be}{\begin{eqnarray}}
\newcommand{\ee}{\end{eqnarray}}
\newcommand{\bma}{\begin{matrix}}
\newcommand{\ema}{\cr\end{matrix}}

\newcommand{\D}{\delta}

\usepackage{placeins}
\newcommand{\no}{\nonumber}
\newcommand{\half}{\frac{1}{2}}

\newcommand{\ZZ}{\mathbb{Z}}

\newcommand{\CC}{\mathbb{C}}
\newcommand{\NN}{\mathbb{N}}

\newcommand{\cF}{\mathcal{F}}
\newcommand{\m}{\mu}
\newcommand{\g}{\gamma}

\usepackage{colortbl}
\usepackage{dsfont}
\newcommand{\chiG}{\widetilde{\chi}}

\def\ie{\begin{equation}\begin{aligned}}
\def\fe{\end{aligned}\end{equation}}

\def\Z{\mathbb{Z}}
\def\cM{\mathcal{M}}
\def\A{\alpha}

\def\Tr{\text{Tr}\,}
\def\SL{{\rm SL}}
\def\GL{{\rm GL}}
\def\PSL{{\rm PSL}}

\definecolor{amethyst}{rgb}{0.5, 0, 0.5}

\begin{document}
\title{Holomorphic modular bootstrap revisited
}
\author[1]{Justin Kaidi}
\author[2]{Ying-Hsuan Lin}
\author[3]{Julio Parra-Martinez}
\affiliation[1]{Simons Center for Geometry and Physics, Stony Brook University,
Stony Brook, NY 11794-3636, USA}
\affiliation[2]{Jefferson Physical Laboratory, Harvard University, Cambridge, MA 02138, USA}
\affiliation[3]{Walter Burke Institute for Theoretical Physics,
California Institute of Technology, Pasadena, CA 91125, USA}

\abstract{
In this work we revisit the ``holomorphic modular bootstrap'', i.e. the classification of rational conformal field theories via an analysis of the modular differential equations  satisfied by their characters.
By making use of the representation theory of ${\rm PSL}(2,\mathbb{Z}_n)$, we describe a method to classify allowed central charges and weights $(c,h_i)$ for theories with any number of characters $d$.
This allows us to avoid various bottlenecks encountered previously in the literature, and leads to a classification of consistent characters up to $d=5$ whose modular differential equations are uniquely fixed in terms of $(c,h_i)$.
In the process, we identify the full set of constraints on the allowed values of the Wronskian index for fixed $d\leq 5$.
}

\maketitle
\section{Introduction}

In the general program of classifying rational conformal field theories (RCFTs) \cite{Friedan:1983xq}, substantial effort has been put into classifying the following proxies:
modular tensor categories (MTCs) \cite{turaev2016quantum,Moore:1989vd} and
 vector-valued modular functions (vvmfs)
\cite{gannon2014theory}.\footnote{The acronym vvmf more commonly means vector-valued modular ``forms'' that allow general weights, but this paper only concerns the weight-zero case, i.e. vector-valued modular functions.}
The two are of course intimately related: the modular S- and T-matrices of an MTC capture, in some appropriate sense, the $\PSL(2, \mathbb{Z})$ transformation properties of RCFT characters, which form a vvmf.\footnote{More precisely, the modular S- and T-matrices of the MTC are generically reduced via an identification of modules related by charge conjugation to give the $\PSL(2, \mathbb{Z})$ representation describing the modular transformations of RCFT characters.  See Section~\ref{Sec:Additional}.}
Each proxy has its limitations.
For MTCs, it is a famous open problem whether every MTC is realized by an RCFT, and even if so infinitely many RCFTs can realize the same MTC.
For vvmfs, having a set of candidate characters with consistent modular transformations does not guarantee that an actual RCFT exists; for example, a famous set of consistent characters/partition functions that are not known to correspond to fully consistent theories is the series of extremal meromorphic CFTs conjectured by Witten \cite{Witten:2007kt}.
As such, the classification of these proxies should be regarded only as an intermediate step towards the classification of RCFTs.

The classification of MTCs has received a significant amount of attention, and has been successfully carried out for MTCs with up to five modules \cite{Rowell:2007dge,bruillard2016classification}. As for vvmfs, there exist several methods to construct them (see \cite{Cheng:2020srs} for a nice comparison). Among these is the theory of vvmfs due to Bantay and Gannon, which describes the entire space of vvmfs transforming under a given $\PSL(2,\Z)$ representation in terms of a single characteristic matrix \cite{bantay2006conformal,bantay2007vector}.
The explicit analytic construction of such a matrix, however, is technically rather involved. A less sophisticated but more practical approach is the method of modular differential equations (MDEs) \cite{Mathur:1988na}, which stems from the observation that every component $\chi_i(\tau)$ of a rank-$d$ vvmf satisfies an order-$d$ MDE,
\begin{equation}
\label{sec:firstMDEdef}
\left[D^{(d)} + \sum_{k=0}^{d-1} \phi_k(\tau) D^{(k)} \right] \chi_i(\tau) = 0~,
\end{equation}
where $D^{(k)}$ is the $k$-th order modular covariant derivative, to be defined below, and the coefficients $\phi_k(\tau)$ are meromorphic modular forms of weight  $2(d-k)$. As will be explained below, we will be able to restrict to meromorphic functions $\phi_k(\tau)$ with certain types of singularities, in which case the space of such modular forms becomes finite-dimensional. This allows one to classify vvmfs by constructing and solving the most general MDE of a given order, e.g. as a Fourier series in $q:=e^{2\pi i \tau}$. This version of the classification problem is sometimes known as the Mathur-Mukhi-Sen program \cite{Mathur:1988na,Mathur:1988gt}, and has been extensively explored in e.g.~\cite{Hampapura:2015cea,Gaberdiel:2016zke,Hampapura:2016mmz,mason2018vertex,Chandra:2018pjq,Mukhi:2020gnj,Das:2020wsi,Bae:2020xzl}. Following \cite{Mukhi:2019xjy}, we will refer to this program as the ``holomorphic modular bootstrap."

An important complication in the MDE approach to classifying vvmfs is that the space of such differential equations can be difficult to navigate. Each MDE has many free coefficients and \textit{a priori} it is not clear what a good organizing principle is. As will be reviewed below, a useful quantity in this respect is the Wronskian index $\ell$, which counts the number of zeros in the Wronskian determinant of the solutions to the MDE (times six)---see Section \ref{sec:Wronskindmonic} for details. A rough classification of MDEs can then be given in terms of the pair $(d, \ell)$.
An even more useful classification principle is the \textit{monodromy} of the solutions of the MDE, which encodes the detailed modular properties of its solutions. For instance, we may study the monodromy of the vvmf $\vec\chi(q)$ around $q=0$ as we traverse a loop $\gamma(s) = e^{2 \pi i s} q$ from $s=0$ to $s=1$. This monodromy describes the action of the modular $T$ transformation on the characters:
\begin{equation}
       \label{eq:Aidefs} \vec\chi( q) \rightarrow T \,\vec\chi(q)\,, \qquad  T = \text{diag} (e^{2\pi i \alpha_1}, \cdots, e^{2\pi i \alpha_d})
\end{equation}
which is specified by a list of exponents $\alpha_i := h_i - \frac{c}{24}$ determined by the parameters in the MDE. These exponents encode the central charge $c$ and weights $h_i$ of the putative RCFT.

As it turns out, the characters of any RCFT are solutions to  an MDE with a \textit{finite} monodromy group. This was first observed empirically by Mathur and Sen in \cite{mathur1989group}, who used this fact to classify two-character theories, as well as a subset of three-character theories. A concrete mathematical conjecture \cite{atkin1971modular}, sometimes known as the {\it integrality conjecture}, was already available at the time to ``explain'' this observation.  This conjecture states that the integrality of Fourier coefficients of a vvmf (with rational exponents) implies that every component is invariant under a principal congruence subgroup of $\PSL(2, \mathbb{Z})$, denoted by $\Gamma(n)$ for some $n\in\mathbb{N}$. Hence, the vvmf transforms in a representation of $\PSL(2, \mathbb{Z})/\Gamma(n) \simeq \PSL(2, \mathbb{Z}_n)$, and the corresponding MDE has finite monodromy as claimed. For general vvmfs, this remains a conjecture, but for RCFT characters, it is a proven fact known as the {\it congruence property} in the MTC literature \cite{Ng:2012ty}. As we will see, integrality and its relation to the monodromy and congruence properties of characters will form the cornerstone of our analysis.

The purpose of this paper is to further the progress on the classification of RCFTs through the holomorphic modular bootstrap. We will see that the classification of exponents $\{\alpha_i\}$, or equivalently of the central charge and weights $(c,h_i)$ of the putative RCFT, reduces to a problem in $\PSL(2,\Z)$ representation theory. In particular, the integrality conjecture allows us to focus on only $\PSL(2,\Z)$ representations $\rho : \PSL(2,\Z) \to \GL(d,\mathbb{C})$ whose kernels are congruence subgroups containing $\Gamma(n)$, which enables us to study a reduced problem about the representation theory of $\PSL(2, \mathbb{Z}_n)$.  We avoid the proliferation of free parameters by focusing on a class of \emph{rigid} MDEs with $d\leq 5$, which are uniquely specified by the exponents $\{\alpha_i\}$. The solutions of such MDEs yield the characters for all theories with $(d,\ell) = (2,0)$, $(2,2)$, $(3,0)$, $(4,0)$, and $(5,0)$. Having imposed integrality, physical requirements such as positivity and the existence of a vacuum will then be imposed by hand when appropriate. For ease of reference, here we summarize the tables in which potential theories of type $(d,\ell)$ are enumerated:

\vspace{10pt}
\begin{tabular}{ccccc}
     $(d, \ell) = (2,0)$
     & $(d, \ell) = (2,2)$
     & $(d, \ell) = (3,0)$ & $(d, \ell) = (4,0)$ & $(d, \ell) = (5,0)$ \\[5pt]
     Table \ref{table:ell=02}
    &   Table \ref{table:ell=02}
     & Table \ref{tab:3dl0tabimprim1} & Table \ref{Tab:Unitary} & Table \ref{tab:5dl0physical}
\end{tabular}
\vspace{10pt}

For non-rigid MDEs, $\PSL(2, \mathbb{Z}_n)$ representation theory still permits a classification of allowed exponents for integral characters, and we carry this classification out. However, in this case imposing physical constraints such as positivity and existence of a vacuum becomes more difficult. This prevents us from obtaining a classification of physical characters. In any case, our analysis allows us to obtain restrictions on the allowed values of the Wronskian index, showing that $\ell \in 2 \ZZ$ for $d=2,4$, while $\ell \in 3\ZZ$ for $d=3$. For $d=5$, there is no constraint on $\ell$.

It should be noted that our classification of vvmfs up to dimension 5 can include RCFTs with up to 9 modules (vacuum and 4 pairs of charge-conjugated modules).  Therefore, it is clear that our results are not contained in the current classification of MTCs up to 5 modules \cite{Rowell:2007dge,bruillard2016classification}.

\subsection*{Outline}
After a brief review of MDEs in Section \ref{sec:MDE}, we use $\PSL(2,\ZZ_n)$ representation theory to constrain the possible exponents for rigid MDEs in Section \ref{sec:mainsec}.
In Section \ref{sec:mainsec2}, we use these allowed exponents as an input for a computerized scan to identify the choices which give physically-sensible RCFTs, i.e. those  satisfying positivity, existence of a vacuum, and other conditions described in Section \ref{Sec:Additional}.
In Section \ref{sec:MathurSen} we give an alternative derivation of our 3- and 4-character results by interpreting the finite monodromy groups as finite subgroups of $\GL(d,\CC)$. Though this is technically harder, it serves as a nontrivial verification of our results, and also makes contact with previous work by Mathur and Sen \cite{mathur1989group}. Since the final results are already presented in  Section \ref{sec:mainsec2}, the busy reader may skip this section.

There are a number of appendices. In Appendix \ref{app:extraexps}, we list the exponents giving rise to integral characters, including those for non-rigid MDEs. This gives a complete classification of exponents for quasicharacters. In Appendix \ref{app:furtherdenom}, we give some additional information about the allowed exponents. In Appendix \ref{app:SLdfinsubgs},  we review the finite subgroups of $\SL(3,\CC)$ and $\SL(4,\CC)$, which are needed for the analysis in Section~\ref{sec:MathurSen}. Finally, Appendix~\ref{app:4dimprim} gives details on the derivation of some exponents quoted in Section \ref{sec:MathurSen}.

\section{Modular differential equation}
\label{sec:MDE}

The characters of an RCFT (or, more generally, the components of a vvmf) form solutions to an ordinary linear differential equation with coefficient functions being modular forms.  This fact follows from the existence of null vectors in the vacuum module \cite{Eguchi:1986sb,zhu1990vertex}, and also from the meromorphy and automorphy of the characters \cite{Anderson:1987ge}.  While the two perspectives produce the same differential equation \cite{Gaberdiel:2008pr}, the latter perspective, often dubbed the Wronskian method, does not require knowing the specifics of the vertex operator algebra (VOA), and hence provides an elegant classification framework for RCFTs, originally proposed in \cite{Mathur:1988na,Mathur:1988gt}.

We briefly sketch the derivation of the modular differential equation (MDE) via the Wronskian method.
The $d$ independent characters $\chi_i(\tau)$ of an RCFT are
meromorphic in the upper half plane away from the cusps,
and automorphic under $\PSL(2,\Z)$ modular transformations.
Note that meromorphy forbids any singularities in the interior of the upper half plane, and hence imposes a growth condition on the characters.
Let $E_{2k}(\tau)$ denote the holomorphic Eisenstein series, and define a covariant derivative
\bea
D_w :=  \frac{1}{2\pi i} \frac{d}{d\tau} - \frac{w}{12} E_2(\tau)
\eea
acting on modular forms of weight $w$. The order-$d$ derivative may be defined in terms of this as \bea
D^{(k)} := \prod_{s=1}^k D_{2s-2}~.
\eea
If $f(\tau)$ is any linear combination of the characters $\chi_i(\tau), \, i=1, \dotsc, d$, then it is clear that
\ie
\det\begin{pmatrix}
f(\tau) & \chi_1(\tau) & \dotsc & \chi_d(\tau)
\\
D^{(1)} f(\tau) & D^{(1)} \chi_1(\tau) & \dotsc & D^{(1)} \chi_d(\tau)
\\
\vdots & \vdots && \vdots
\\
D^{(d)} f(\tau) & D^{(d)} \chi_1(\tau) & \dotsc & D^{(d)} \chi_d(\tau)
\end{pmatrix} = 0 \, .
\fe
Laplace expanding the above determinant gives the MDE
\ie
D^{(d)} f(\tau) + \sum_{k=0}^{d-1} \phi_k(\tau) D^{(k)} f(\tau) = 0 \, ,
\fe
where
\ie
\label{Wronskian}
\phi_k(\tau) = (-)^{n-k} \frac{W_k(\tau)}{W(\tau)} \, ,
\quad
W_k(\tau) = \det\begin{pmatrix}
\chi_1(\tau) & \dotsc & \chi_d(\tau)
\\
D^{(1)} \chi_1(\tau) & \dotsc & D^{(1)} \chi_d(\tau)
\\
\vdots & \vdots & \vdots
\\
D^{(k-1)}_\tau \chi_1(\tau) & \dotsc & D^{(k-1)}_\tau \chi_d(\tau)
\\
D^{(k+1)}_\tau \chi_1(\tau) & \dotsc & D^{(k+1)}_\tau \chi_d(\tau)
\\
\vdots & \vdots & \vdots
\\
D^{(d)} \chi_1(\tau) & \dotsc & D^{(d)} \chi_d(\tau)
\end{pmatrix} \, .
\fe
In particular, $W(\tau) := W_d(\tau)$ is called the Wronskian. Such differential equations easily yield power series solutions in $q=e^{2\pi i \tau}$ (i.e. Fourier series in $\tau$),
\bea
\chi_i(\tau) = q^{\alpha_i} ( m_{i,0} +  m_{i,1}\, q + m_{i,2}\,q^{2} + \cdots)\,,
\label{eq:fseriessol}
\eea
where we have defined the set of exponents $\{\A_i\}$, which for vvmfs interpretable as character vectors of RCFTs are given by
\be
\alpha_i = h_i - \frac{c}{24}\,,
\ee
with $c$ the central charge and $h_i$ the chiral dimensions of the characters.

Although the coefficient functions $\phi_k(\tau)$ are left undetermined from the derivation above,
we do know that they are meromorphic modular forms of weight $2(d-k)$, and the space of meromorphic modular forms consists of all rational expressions in $E_4(\tau)$ and $E_6(\tau)$.  If certain restrictions are imposed on the singularities of $\phi_k(\tau)$, then the restricted subspace is in fact finite-dimensional.  This finiteness is what makes the MDE classification of RCFTs highly effective and constraining.  Our next subject is a scheme to impose such restrictions on the singularities.

\subsection{Wronskian index and monicity}
\label{sec:Wronskindmonic}
In organizing MDEs, it is useful to introduce the \textit{Wronskian index} $\ell$, defined to be six times the number of zeros of the Wronskian $W(\tau)$ in the fundamental domain $\cF$ of $\PSL(2,\Z)$.  If a zero is located at the boundary of $\cF$, then it is weighted by the conical fraction, namely $\frac13$ at the $ST$-invariant point $\omega = e^{\frac{2\pi i}3}$ and $\frac12$ at the $S$-invariant point $i$,
\bea
\label{eq:defl}
\ell : = 6 \left( \half \mathrm{ord}_i(W) + {1\over 3} \mathrm{ord}_\omega(W) + \sum_{\substack{p \in \cF/ \SL(2, \ZZ) \\ p \neq i, \omega}}  \mathrm{ord}_p(W)\right)~.
\eea
Via \eqref{Wronskian}, information about $\ell$, and hence about the zeros of $W(\tau)$, imposes constraints on the allowed singularities of the coefficient functions $\phi_k(\tau)$. For instance, consider the case of $(d,\ell)=(2,2)$. In this case (\ref{eq:defl}) implies that we have a single zero of $W(\tau)$ at $\tau = \omega$. Recalling that $E_4(e^{2\pi i / 3})=0$ and $E_6(i)=0$,  the most general  MDE with these labels is of the form
\bea
\label{eq:2dMDE}
\left[D^{(2)}  +  \m_1 {E_6 \over E_4} D^{(1)} + \g E_4 \right] f(\tau) \,= \,0\,.
\eea
Likewise for $(d, \ell)=(3,3)$ the most general MDE is
\bea
\label{eq:3dMDE}
\left[ D^{(3)} \! +\m_1 {E_4^2 \over E_6} D^{(2)} + \g_1 E_4 D^{(1)} + \m_2 {E_4^3 \over E_6}+\g_2 E_6\right]\! f(\tau)\,= \, 0~.
\eea
 Note that for $\ell < 6$, the denominators of the coefficient functions are always just powers of $E_4(\tau)$ or $E_6(\tau)$.

 Due to meromorphy, $\ell$ is also related to the pole order at the cusp, and hence to the eigenvalues $\A_i$ of $T$ defined in (\ref{eq:Aidefs}). This follows from the \textit{valence formula}, which for meromorphic modular forms $f(\tau)$ of weight $w$ reads \cite{diamond2005first}
\bea
\sum_{p \in \cF/ \SL(2,\ZZ)} \mathrm{ord}_p(f) = {w \over 12}
\eea
with $\mathrm{ord}_p(f)$ counting the order of zeros/poles of $f(\tau)$ at $\tau = p$ (counted positively for zeros and negatively for poles). This together with the fact that $W(\tau)$ has weight $d(d-1)$ gives the following relation between $\ell$ and the $\A_i$,
\ie
\frac{\ell}{6} = \frac{d(d-1)}{12} - \sum_{i=1}^d \A_i \, , \hspace{0.4 in} \A_i = h_i - \frac{c}{24} \, .
\fe
In particular, we see that
\ie
\label{WronskianDetT}
\ell = \frac{d(d-1)}{2} - \frac{\log \det T}{2\pi i} \,\,\,\,\mod 6 ~ .
\fe

In \cite{Mathur:1988na,Mathur:1988gt}, Mathur, Mukhi, and Sen proposed a classification scheme for RCFTs in which one progresses not only in increasing $d$, but also increasing $\ell$.
Let us provide some intuition for why it is reasonable to focus on smaller values of $\ell$, or even restrict to $\ell = 0$.  First, any set of characters can be multiplied by $j(\tau)^{\frac{m}{3}}$ with $m$ a natural number to produce another set of characters, corresponding physically to taking the tensor product with $m$ copies of the $(E_8)_1$ WZW model.  Because $j(\tau)$ has a simple zero at the $ST$-invariant point, this extra factor increases $\ell$ by $2d$.  One may want to ignore such trivial tensor products by stripping off overall factors of $j(\tau)^{\frac{1}{3}}$ whenever possible to minimize $\ell$.  For $\ell < 2d$, one always gets a set of characters without any removable $j(\tau)^{\frac{m}{3}}$ factor.

An RCFT or its set of characters is said to be {\it monic} if the Wronskian index vanishes $\ell = 0$.  Equivalently, an MDE is called {\it monic} if all the coefficient functions are polynomial in $E_4(\tau)$ and $E_6(\tau)$.  Since $E_4(\tau)$ and $E_6(\tau)$ are holomorphic in the upper half-plane, this means that the coefficient functions are regular.  Monic MDEs have been a subject of intensive studies for several reasons.  First, many of the familiar classes of theories, including all Virasoro minimal models and many WZW models, are monic; generally, since the monic restriction makes the MDEs maximally constraining, the solutions are easier to classify.  Second, a non-monic MDE can sometimes be monicized by composing with an additional modular differential operator, at the cost of increasing the order of the MDE and introducing fictitious solutions.  For instance, the unique character $j(\tau)^{\frac13}$ of the $(E_8)_1$ WZW model is non-monic (in particular $\ell = 2$), but it also solves a second order monic MDE.  While  it is generally unclear which non-monic MDEs are monicizable, what is clear is that monic MDEs also allow for the discovery of non-monic RCFTs.  Finally, there is an interesting conjecture \cite{Beem:2017ooy} that the Schur index of a four-dimensional $\mathcal{N} = 2$ SCFT always solves a monic MDE of a definite order.\footnote{The two-dimensional vertex operator algebra is not necessarily monic, but the MDE is conjectured to be monicizable.
}

\subsection{Rigid modular differential equations}

Having introduced the most general MDE, we now restrict to a special class of MDEs which we refer to as \textit{rigid}. An MDE is said to be rigid if the coefficient functions cannot be  adjusted without changing the exponents $\alpha_i=h_i - \frac{c}{24}$ or introducing extra singularities. We will see that almost all rigid MDEs are monic, but not all monic MDEs are rigid. Indeed, it is a simple combinatoric exercise to enumerate the free parameters of monic MDEs of order $d$: since there is no (weakly) holomorphic modular form of weight 2, the $D^{(d-1)}$ term must be absent, and this fixes the sum of exponents to be $\sum_i \A_i = \frac{d(d-1)}{12}$.  Up to $d = 5$, there is exactly one free parameter for each of the lower order derivatives, and hence the exponents uniquely fix the monic MDE.  By contrast, for $d \geq 6$ there are extra free parameters, and thus monic MDEs for $d \ge 6$ are not rigid.

To be more explicit, the relations between the exponents and the MDE parameters for monic $d=2$ and $d=3$ are as follows.  For $d=2$, the most general monic MDE takes the form
\begin{equation}
    \left[ D^{(2)} + \gamma E_4(\tau) \right] \chi_i(\tau) = 0 \, ,
\end{equation}
as can be obtained from (\ref{eq:2dMDE}) by taking $\m_1 = 0$. In this case the single coefficient is determined in terms of the exponents by
\begin{equation}
    \gamma = \alpha_1 \alpha_2 \, .
\end{equation}
For $d=3$, the most general monic MDE takes the form
\begin{equation}
    \left[ D^{(3)} + \gamma_1 E_4(\tau) D^{(1)} + \gamma_2 E_6(\tau) \right] \chi_i(\tau) = 0 \, ,
\end{equation}
as can be obtained from (\ref{eq:3dMDE}) by taking $\m_1 =\m_2 = 0$. In this case the two undetermined coefficients are related to the exponents by
\begin{equation}
    \gamma_1 = \alpha_1 \alpha_2 + \alpha_2 \alpha_3 + \alpha_3 \alpha_1 - \frac{1}{18} \, , \quad \gamma_2 = - \alpha_1 \alpha_2 \alpha_3 \, .
\end{equation}
Similar results are easily obtained for monic $d=4,5$.

Finally, there is a single family of rigid MDEs which are non-monic---namely $d = 2$ MDEs with Wronskian index $\ell = 2$. As was shown in (\ref{eq:2dMDE}), this MDE has two free parameters, which can be fixed in terms of the two exponents.

The aim of this paper is to classify character solutions to rigid MDEs.  As we will see, modular representation theory will constrain the exponents $\A_i := h_i - \frac{c}{24}$ to lie within finite sets (modulo integer shifts), and since the exponents completely fix the MDE, we may analyze all possible MDEs by scanning over a discrete lattice of possibilities.  By contrast, without rigidity, the space of MDEs and solutions involves parameters that are continuous before one imposes the integrality of the Fourier coefficients.

\subsection{Physical characters and quasicharacters}
\label{Sec:Additional}

As is by now familiar, any $d$-dimensional vvmf solves an order-$d$ MDE. Conversely, the solutions to any MDE form a vvmf.
However, not all MDEs give rise to vvmfs which are interpretable as character vectors for some RCFT. Further requirements must be imposed to obtain physical characters.
We find the following constraints well-motivated:
\begin{enumerate}
\item{\bf Weak holomorphy:} The solutions must be holomorphic (regular) away from the cusps.

\item{\bf Integrality:} The solutions to the MDE must admit a normalization in which all Fourier coefficients are integers, to be interpreted as state degeneracies.

\item  {\bf Vacuum:} At least one solution $\chi_i(\tau)$
under suitable normalization has
unit leading Fourier coefficient and unit modular pairing coefficient, to be interpreted as the vacuum.  More details are given below.

\item{\bf Positivity:} The Fourier coefficients of the solutions should be purely positive. To limit ourselves to a smaller number of solutions, we also require that $c>0$ and $h_i \ge 0$ for at least one interpretation of the vacuum.  These conditions would be required of unitary theories,
but we will also discover some non-unitary theories satisfying the above criteria.\footnote{There are of course other conditions we could impose if we were to strictly enforce unitarity, such as $c\ge1$ except for minimal models, or requiring the modular S-matrix to satisfy $S_{0i} > 0$ for all $i$.}

\end{enumerate}

We now provide more comments on some of these constraints. First, note that in the modular bootstrap program, one of the most difficult consistency conditions to impose is the integrality of Fourier coefficients.  It is remarkable that in the rational world, the theory of vvmfs draws direct connections between integrality and nontrivial physical statements, as was recently explored in \cite{Kaidi:2020ecu}. As already noted in the introduction, in the present work we will  guarantee integrality by requiring that the modular S- and T-matrices describing the modular properties of the vvmf furnish a representation of $\PSL(2,\ZZ_n)$, {\`a} la the integrality conjecture.

Vvmfs satisfying weak holomorphy and integrality, but not necessarily vacuum and positivity are called \emph{quasicharacters} \cite{Chandra:2018pjq,Mukhi:2020gnj}.
Depending on whether we choose to impose the latter two conditions, our work can be taken to classify either quasicharacters or physical characters. As we explain in Section~\ref{sec:mainsec}, for solutions to monic MDEs, weak holomorphy and integrality are automatic once the consistent exponents are classified. On the other hand, when appropriate we will check vacuum and positivity by explicitly solving the MDEs, as we are not aware of any alternative.

Next, we give details on the vacuum condition. To identify whether a vvmf admits a legitimate vacuum character, we must check that under suitable normalization, at least one solution has unit leading Fourier coefficient. This is done as follows. One begins by stripping off the greatest common factor of all Fourier coefficients of the vvmf, and then solving modular invariance
to compute a ``bare'' modular pairing matrix $M'_{ij} = m'_i \D_{ij}$, such that the combination $\sum_{i,j} M'_{ij} \chi_i(\tau) \overline{\chi_j}(\overline{\tau})$ is modular invariant. This modular pairing matrix does not generically have integer entries. To make it integer, we may do the following. For any $i$, if $m'_i$ is rational and takes the form $m'_i = m_i (p_i/q_i)^2$ for integer $m_i, p_i, q_i$ with $(p_i,q_i)=1$, then we can renormalize $\chi_i(\tau) \to p_i \chi_i(\tau)$, $\chi_j(\tau) \to q_i \chi_j(\tau)$ for $j \neq i$ to obtain a new $M'_{ij}$.
Iterating over $i$, where we always choose the triplet $(m_i, p_i, q_i)$ that minimizes $q_i$, we eventually obtain a final modular pairing matrix $M_{ij} = m_i \D_{ij}$ with integral degeneracies.
It is then in this normalization that we look for a character with unit leading Fourier coefficient.

We illustrate this procedure with the following example.  Consider the solutions to a third order monic MDE under exponents $(-\frac{11}{30}, -\frac16, \frac{31}{30})$
\begin{equation}
\begin{array}{c}
 \chi_1(\tau) = q^{-\frac{11}{30}}(1 + 253 q + 4642 q^2 + 43824 q^3 + \mathcal{O}(q^4)) \, , \\
  \chi_2(\tau) = q^{-\frac16}(11 + 2464 q + 43614 q^2 + 393250 q^3 + \mathcal{O}(q^4)) \, , \\
 \chi_3(\tau) = q^{\frac{31}{30}} \left(242 + 4092 q + 35123 q^2 + 221464 q^3 + \mathcal{O}(q^4) \right) \, . \\
\end{array}
\end{equation}
The character $\chi_1(\tau)$ may naively seem to give a legitimate vacuum character since it has unit leading Fourier coefficient. However, this is not correct. Indeed, in the current normalization the ``bare'' modular pairing matrix is found to be
\begin{equation}
M' =
\left( \renewcommand*{\arraystretch}{0.8} \begin{array}{ccc}
1&&
\\
& \frac{1}{50}&
\\
& & 1
\end{array} \right)\, .
\end{equation}
which is not integral. To make it integral, we note that $\frac{1}{50} = \frac{2}{10^2}$, i.e. $(m_2, p_2, q_2) = (2, 1, 10)$ in our previous notation. We must then rescale the characters by factors of $(10, 1, 10)$ to obtain
\begin{align}
 \chi_1(\tau) &= 10 q^{-\frac{11}{30}}(1 + 253 q + 4642 q^2 + 43824 q^3 + \mathcal{O}(q^4)) \, , \nonumber \\
  \chi_2(\tau) &= q^{-\frac16}(11 + 2464 q + 43614 q^2 + 393250 q^3 + \mathcal{O}(q^4)) \, , \\
 \chi_3(\tau) &= 10 q^{\frac{31}{30}} \left(242 + 4092 q + 35123 q^2 + 221464 q^3 + \mathcal{O}(q^4) \right) \nonumber \, ,
\end{align}
which has an integral modular pairing matrix
\begin{equation}
M =
\left( \renewcommand*{\arraystretch}{0.8} \begin{array}{ccc}
1&&
\\
&2&
\\
& & 1
\end{array} \right)\, .
\end{equation}
In this normalization, we see that there is no candidate vacuum character.

As a final comment, note that it is possible for multiple characters to be interpretable as the vacuum. For example, consider the following pair of characters, which provide an integral set of solutions to a second-order MDE with exponents $\{-{1\over 60},{11\over 60}\}$,
\bea
\chi_1(q) &=& q^{-{1 \over 60}}(1 + q + q^2 + q^3  + \mathcal{O}(q^4))~,
\no\\
\chi_2(q) &=& q^{11\over 60}\,(1 + q^2 + q^3 + \mathcal{O}(q^4))~.
\eea
Since the bare pairing matrix is already integral $M' = {\rm diag}(1,1)$, and both characters have unit leading Fourier coefficient, we see that either one of them can be interpreted as the vacuum character. Interpreting $\chi_1(q)$ as the vacuum character leads to a tentative theory with $(c,h) = ({2\over 5}, {1\over5})$, while interpreting $\chi_2(q)$ as the vacuum leads to a tentative theory with $(c,h) = (-{22\over 5}, - {1\over5})$. We recognize the latter as the $\cM_{2,5}$ minimal model.
Several other concrete examples will be encountered in the following sections.

\paragraph{MTC Structure?} Above we listed four physical conditions which will be imposed on characters of our tentative theories. One might be tempted to further impose that S- and T-matrices describing the modular transformations of the vvmfs furnish the modular data of an MTC. For instance once could require that $S$, when inserted in the Verlinde formula \cite{Verlinde:1988sn},
\bea
N_{ijk} = \sum_\ell {S_{i\ell} S_{j \ell} S_{k \ell}^* \over S_{0 \ell}}~,
\eea
yields positive integer fusion coefficients.
However, this is incorrect. Indeed, the S- and T-matrices of an MTC encode the modular properties of \emph{refined} characters---those manifesting not just $q = \exp(2\pi i \tau)$, but also the chemical potentials for a set of commuting conserved charges, such as the $G$ quantum numbers for a $\widehat G_k$ WZW model, or, more generally, commuting quantum KdV charges.
The classification of RCFTs through vvmfs ignores such extra chemical potentials,\footnote{See however \cite{Pan:2021ulr} for recent work on flavored MDEs.}
and the $\PSL(2, \mathbb{Z})$ transformation properties of the characters are instead captured by {\it reduced} S- and T-matrices that need not fit inside the structure of an MTC.  In particular, the Verlinde formula applied to reduced S-matrices generally does not generate non-negative integral fusion coefficients.

As a concrete example, the $\widehat{\mathfrak{so}(2r)_1}$ WZW model has four modules with four distinct {\it refined} characters
\ie
& \chi_{\widehat{\omega}_0}(\tau, z) = \frac12 \frac{\theta_3^r(\tau, z) + \theta_4^r(\tau, z)}{\eta^r(\tau)} \, ,
\quad
\chi_{\widehat{\omega}_1}(\tau, z) = \frac12 \frac{\theta_3^r(\tau, z) - \theta_4^r(\tau, z)}{\eta^r(\tau)} \, ,
\\
& \chi_{\widehat{\omega}_{r-1}}(\tau, z) = \frac12 \frac{\theta_2^r(\tau, z) + \theta_1^r(\tau, z)}{\eta^r(\tau)} \, ,
\quad
\chi_{\widehat{\omega}_r}(\tau, z) = \frac12 \frac{\theta_2^r(\tau, z) - \theta_1^r(\tau, z)}{\eta^r(\tau)} \, ,
\fe
where $z$ is the chemical potential for the U(1) global symmetry that simultaneously rotates the $r$ pairs of complex fermions by a phase.\footnote{One could turn on $r$ chemical potentials $z_i$ for the $r$ copies of U(1) rotations, but setting $z_i = z$ for all $i$ suffices to distinguish the four modules.
}
For $r = 3$, the refined characters transform according to the $4 \times 4$ modular S- and T-matrices
\ie
\label{ModularST}
S = \frac12
\begin{pmatrix}
1 & 1 & 1 & 1
\\
1 & 1 & -1 & -1
\\
1 & -1 & i & -i
\\
1 & -1 & -i & i
\end{pmatrix}
\, ,
\quad
T =
\begin{pmatrix}
e^{-\frac{\pi i}4} & 0 & 0 & 0
\\
0 & e^{\frac{3\pi i}4} & 0 & 0
\\
0 & 0 & e^{\frac{\pi i}2} & 0
\\
0 & 0 & 0 & e^{\frac{\pi i}2}
\end{pmatrix} \, ,
\fe
When the chemical potential is turned off, two of the refined characters become degenerate, $\chi_{\widehat\omega_2}(\tau, 0) = \chi_{\widehat\omega_3}(\tau, 0)$, and the {\it unrefined} characters transform according to $3 \times 3$ reduced S- and T-matrices
\ie
\label{ReducedST}
 S = \frac12
\begin{pmatrix}
1 & 1 & 1
\\
1 & 1 & -1
\\
2 & -2 & 0
\end{pmatrix}
\, ,
\quad
 T =
\begin{pmatrix}
e^{-\frac{\pi i}4} & 0 & 0
\\
0 & e^{\frac{3\pi i}4} & 0
\\
0 & 0 & e^{\frac{\pi i}2}
\end{pmatrix} \, ,
\fe
Whereas the modular S-matrix \eqref{ModularST} produces non-negative integral fusion coefficients when applying the Verlinde formula, the reduced S-matrix \eqref{ReducedST} does not, and gives for instance
\ie
\widehat \omega_0 \times \widehat \omega_0 = \frac34 \widehat\omega_0 + \frac14 \widehat\omega_1 \, .
\fe
The lesson here is that for generic RCFTs, it does not make sense to demand that the {\it unrefined} characters transform under S- and T-matrices furnishing an MTC structure.

A restricted subset of RCFTs have the same number of modules as of distinct unrefined characters.  In this case, the unrefined characters
\textit{do} transform under modular S- and T-matrices compatible with the structure of an MTC; in particular, the Verlinde formula generates non-negative integer fusion coefficients.
This version of the classification problem
has been extensively studied  with particular focus on $d=2$ \cite{mason2018vertex,grady2020classification} and $d=3$ \cite{franc2020classification}.  However, notice that the aforementioned $\widehat{\mathfrak{so}(2r)_1}$ WZW models, as well as numerous other RCFTs, do not fall under its purview.

This paper adopts the more general approach of classifying characters through the study of MDEs without necessarily demanding an MTC structure.
In the rest of this paper, the word ``character'' refers to unrefined ones, i.e. without extra fugacities, in which case distinct modules of the vertex operator algebra may share identical characters.

\section{Exponents from modular representation theory}
\label{sec:mainsec}
The characters of an RCFT form a vvmf transforming in a representation $\rho$ of $\PSL(2,\Z)$.
At the heart of our approach is the so-called \textit{integrality conjecture} in the theory of vvmfs \cite{bantay2007vector}, or alternatively the congruence property in the modular tensor category (MTC) literature  \cite{Ng:2012ty}. The basic statement is that integrality of the RCFT characters implies that the characters are modular functions for a principal congruence subgroup $\Gamma(n) < \PSL(2,\Z)$. Here $n$ is the order of $T$, or equivalently the least common denominator of the set of exponents $\{h_j - \frac{c}{24}\}$.  It follows that $\rho$ is a representation of $\PSL(2,\Z) / \Gamma(n) \cong \PSL(2,\Z_n)$.

All representations of $\SL(2,\ZZ_n)$ were classified in \cite{nobs1976irreduziblen,nobs1976irreduziblen2,eholzer1995classification}, and it is this result which provides the first step for our classification approach.\footnote{The original works \cite{nobs1976irreduziblen,nobs1976irreduziblen2} are written in German; \cite{eholzer1995classification} offers a nice review in English.  While $\SL(2,\ZZ_n)$ representation theory has been key to the classification of MTCs, it does not appear to have been leveraged fully in the study of MDEs.
}
In this section, we will discuss how to extend the known representation theory of $\SL(2,\ZZ_n)$ to the representation theory of $\PSL(2,\ZZ_n)$, and then use these results to make statements about the possible representations $\rho$ for integral vvmfs. For example, suppose that we fix a particular $n=n_0$ and find that  $\PSL(2,\ZZ_{n_0})$ admits irreducible representations of dimensions $\{d_i^{(n_0)}\}$. If a certain dimension $d=d_0$ were absent from this list, then we would conclude that no $d_0$-character theory admits a T-matrix of order $n_0$. Conversely, fixing a specific dimension $d=d_0$, we can ask for the set of all $\{n_i^{(d_0)}\}$ such that $\PSL(2,\ZZ_{n_i^{(d_0)}})$ \textit{does} admit an irreducible representation of dimension $d_0$. This will constrain the possible orders of the T-matrix, and hence the possible denominators for the exponents $\{h_j - \frac{c}{24}\}$ of RCFTs with $d_0$ characters. These constraints are obtained in Section \ref{sec:denoms};  for any $d_0$, we will see that only a finite number of denominators are allowed. For example, we will find that for 5-character theories, the only allowed denominators are $\{ 5, 10, 11, 15, 22, 30, 33, 66 \}$.

Furthermore, thanks to the existence of computational tools such as GAP \cite{GAP4}, one can obtain not just the lists of possible denominators, but the lists of all possible exponents themselves (mod 1). The basic idea here is a thorough examination of the character table for $\PSL(2,\ZZ_n)$, as will be described in Section \ref{sec:numerators}.  In other words, for fixed $d_0$ we will obtain a finite list of exponents $\{h_j - \frac{c}{24}\}$ mod 1 which can be realized by any $d_0$-dimensional RCFT. Of course, at this stage we are imposing only integrality, so it is not clear which (if any) of the elements of these lists actually correspond to positive characters with a vacuum. However, once the finite list is in hand, it becomes possible to run an explicit computerized scan to implement positivity and existence of a vacuum, and the results of such a scan is given in Section \ref{sec:computerscan}.

Depending on the specific context, one may or may not wish to restrict to irreducible representations of $\PSL(2,\Z_n)$.  For example, an indecomposable MTC can have either a reducible or irreducible representation, so in the classification of MTCs one needs to consider both.  However, in the context of modular differential equations (MDE), it is reasonable to restrict to irreducible representations because for solutions to an MDE transforming under a reducible representation, there exist lower-order MDEs realizing these characters.  In any case, one may regard irreducibility as an assumption of the classification problem pursued in this paper.

\subsection{Representation theory of $\PSL(2,\Z_n)$}
\label{sec:denoms}

As we have just described, the starting point for our classification is the classification of irreducible linear representations of $\SL(2,\Z_n)$ \cite{nobs1976irreduziblen,nobs1976irreduziblen2,eholzer1995classification}.  To begin, note that if $n = \prod_i p_i^{\lambda_i}$ is the prime decomposition of $n$, then
\bea
\SL(2, \Z_n) \cong \prod_i \SL(2, \Z_{p_i^{\lambda_i}}) \, .
\eea
  Thus any irreducible representation of $\SL(2,\Z_n)$ for general $n$ is a tensor product of irreducible representations of $\SL(2, \Z_{p^\lambda})$ for $\lambda \in \mathbb{N}$ and $p$ prime, and it suffices to classify representations of the latter.

  The center of $\SL(2,\Z_{p^\lambda})$ for $p^\lambda>2$ consists of elements $\pm I$, and the quotient by the center gives  $\PSL(2,\Z_{p^\lambda})$.  Schur's lemma states that any element of the center of a group acts as a scalar operator on an irreducible representation, so $-I$ acts by a sign on each irreducible representation of $\SL(2,\Z_{p^\lambda})$.
   Let us define the signature $\sigma$ of an irreducible linear representation of $\SL(2,\Z_{p^\lambda})$ to be $\sigma =\pm$ if $-I$ acts by $\pm 1$, and the signature is multiplicative under tensor product.  The dimensions of irreducible representations of $\SL(2,\Z_{p^\lambda})$ that are reducible under $\SL(2,\Z_{p^\lambda}) \to \SL(2,\Z_{p^{\lambda-1}})$, together with their possible signatures, are summarized in Table~\ref{Tab:SL2ZNRep}.\footnote{If a representation $\rho$ remains irreducible under $\SL(2,\Z_{p^\lambda}) \to \SL(2,\Z_{p^{\lambda-1}})$, then we should regard $\rho$ as a representation of the latter, hence $n$ is reduced by a factor of $p$.
} The irreducible representations of $\PSL(2,\Z_n)$ are those of $\SL(2,\Z_n)$ with positive overall signature $\sigma = +$.
\begin{table}[htp!]
\centering
$
\begin{array}{|c|c|c|c|}
\hline
p & \lambda & \text{irrep dimensions} & \text{signature}
\\\hline
2 & 1 & \{ 1^+, 2^+ \} &
\\
& 2 & \{ 1^-, 2^-, 3^\pm \}  &
\\
& 3 & \{ 2, 3, 4, 6 \} & \pm
\\
& 4 & \{ 3, 6, 8, 12, 24 \} & \pm
\\
& 5 & \{ 6, 12, 16, 24, 48 \} & \pm
\\
& > 5 & \{ 2^{\lambda-1} \} \cup \{ 3 \cdot 2^{\lambda-i} \mid i = 1, 2, 3, 4 \} & \pm
\\\hline
3 & 1 & \{ 1^+, 2^-, 3^+ \} &
\\
\ge 5 & 1 & \frac12(p-1) & (-)^{\frac{p+1}{2}}
\\
&& \frac12(p+1) & (-)^{\frac{p-1}{2}}
\\
&& p-1 & \pm
\\
&& p & +
\\
&& p+1 & \pm \text{ but } - \text{ for } p=5
\\\hline
\ge 3 & \ge 2 & \frac12(p^2-1)p^{\lambda-2} & \pm
\\
&& (p-1)p^{\lambda-1} & \pm
\\
&& (p+1)p^{\lambda-1} & \pm
\\\hline
\end{array}
$
\caption{The dimensions and signatures of irreducible representations of $\SL(2,\Z_{p^\lambda})$ that are reducible under $\SL(2,\Z_{p^\lambda}) \to \SL(2,\Z_{p^{\lambda-1}})$.
}
\label{Tab:SL2ZNRep}
\end{table}

For example, consider the case of $n = 15$.  The representations of $\SL(2,\Z_{15})=\SL(2,\Z_3)\times \SL(2,\Z_5)$ are just tensor products of those of $\SL(2,\Z_3)$ and $\SL(2,\Z_5)$, and from Table \ref{Tab:SL2ZNRep} we find
\ie
& \{ 1^+, 2^-, 3^+ \} \otimes \{ 2^-, 3^+, 4^\pm, 5^+, 6^- \}
\\
&
= \{ 2^-, 3^+, 4^\pm, 5^+, 6^-, 8^\pm, 9^+, 10^-, 12^\pm, 15^+, 18^- \} \, .
\fe
Among them, the irreducible representations of $\PSL(2,\Z_{15})$ are the ones with $\sigma = +$, so the dimensions of the possible irreps of $\PSL(2,\Z_{15})$ are given by
\ie
\{ 3, 4, 5, 8, 9, 12, 15 \} \, .
\fe
This means that only RCFTs with $d$ belonging to this list can have exponents $\{h_j - \frac{c}{24}\}$ with least common denominator 15. For instance, there can be no two-character theory with exponents $\{ {1\over 3}, {1\over 5}\}$ mod 1.

\subsection{Denominator-rank constraints}

Our goal is now to obtain the set of all possible denominators for fixed number $d \geq 2$ of characters.
Let us denote this set of denominators by $\mathfrak{den}(d)$; in other words, this is the set of $n \in \mathbb{N}$ such that $\PSL(2,\Z_n)$ has a $d$-dimensional irreducible representation (that is reducible under $\SL(2,\Z_n) \to \SL(2,\Z_k)$ for all $k | n$).

To obtain this set, we begin by defining an intermediate set $\mathfrak{den}_0(d)$, which is the set of $p^\lambda$ such that $\SL(2,\Z_{p^\lambda})$ has a $d$-dimensional irreducible representation (that is reducible under $\SL(2,\Z_{p^\lambda}) \to \SL(2,\Z_{p^{\lambda-1}})$).  Moreover, $\mathfrak{den}_0(d)$ retains the information about the signatures.  Explicitly, we see that
\ie
\label{den0}
&\mathfrak{den}_0(1) = \{ 1^+, 2^+, 3^+, 4^-, 6^+, 12^- \} \, ,
\\
& \mathfrak{den}_0(2) = \{ 2^+, 3^-, 4^-, 5^-, 8^\pm \} \, ,
\\
& \mathfrak{den}_0(3) = \{ 3^+, 4^\pm, 5^+, 7^+, 8^\pm, 16^\pm \} \, ,
\\
& \mathfrak{den}_0(4) = \{ 5^\pm, 7^-, 8^\pm, 9^\pm \} \, ,
\\
& \mathfrak{den}_0(5) = \{ 5^+, 11^+ \} \, ,
\fe
In terms of these we can then express $\mathfrak{den}(d)$ as follows. First, define the set $\cF$ of all possible factorizations $f$ of $d$ into integers $\delta_i^{(f)}\geq 2$ (not necessarily prime). For example, for $d=4$ we have $\cF = \left\{f_1, f_2\right\} = \left\{ \{4\}, \{2,2\}\right\}$ and $\delta_1^{(1)}=4$, $\delta_1^{(2)}=\delta_2^{(2)}=2$. Then we may write
\ie
\mathfrak{den}(d) =P_+ \, \bigcup_{f\in \cF}  \mathrm{lcm}\left[ \mathfrak{den}_0(1) ,  \mathfrak{den}_0(\delta_1^{(f)}), \dots,  \mathfrak{den}_0(\delta_{|f|}^{(f)}) \right] \, ,
\fe
where lcm denotes the least common multiple and $P_+$ means that we project to those representations with overall signature $\sigma = +$. We are then able to obtain the following complete lists of acceptable denominators for RCFTs with $d=1, \dots, 5$,
\ie
\label{den}
\mathfrak{den}(1) &= \{ 1, 2, 3, 6 \} \, ,
\\
\mathfrak{den}(2)
&= \{ 2, 4, 6, 8, 12, 20, 24, 60 \} \, ,
\\
\mathfrak{den}(3)
&= \{ 3, 4, 5, 6, 7, 8, 10, 12, 14, 15, 16, 21, 24, 30, 42, 48 \} \, ,
\\
\mathfrak{den}(4) &= \{ 2, 3, 4, 5, 6, 8, 9, 10, 12, 15, 18, 20, 24, 28, 30, 36, 40, 60, 84, 120 \} \, ,
\\
\mathfrak{den}(5)
&= \{ 5, 10, 11, 15, 22, 30, 33, 66 \} \, .
\fe
Of course, there is no obstruction to proceeding to higher $d$. We reemphasize that $\mathfrak{den}(d)$ is a finite set for any given $d$.

\subsection{Exponents from character tables}
\label{sec:numerators}
We have now seen how to restrict the set of possible \textit{denominators} of the exponents $\{h_j - \frac{c}{24}\}$ to a finite list. This immediately implies a finite list of exponents $\{h_j - \frac{c}{24}\}$ mod 1, obtained by simply listing all possible numerators less than each denominator. However, we will now see that one can restrict the space of allowed exponents even further.

\begin{table}[!t]
\centering
$
\begin{array}{c|ccccccc}
 &1a & 6 a & 6 b & 2 a & 3 a & 3 b & 4 a \\\hline
\chiG_1 & 1 & 1 & 1 & 1 & 1 & 1 & 1 \\
 \chiG_2 &1 & \bar{\omega } & \omega  & 1 & \omega  & \bar{\omega } & 1 \\
\chiG_3& 1 & \omega  & \bar{\omega } & 1 & \bar{\omega } & \omega  & 1 \\
\chiG_4 & 2 & 1 & 1 & -2 & -1 & -1 & 0 \\
\chiG_5 & 2 & \omega  & \bar{\omega } & -2 & -\bar{\omega } & -\omega  & 0 \\
\chiG_6 & 2 & \bar{\omega } & \omega  & -2 & -\omega  & -\bar{\omega } & 0 \\
\chiG_7 & 3 & 0 & 0 & 3 & 0 & 0 & -1 \\
\end{array}
$
\caption{Character table of $\SL(2,\Z_3)$. $T$ is in conjugacy class 3$b$, while $T^2$ is in the conjugacy class 3$a$.}
\label{Tab:SL23}
\end{table}

The key tool for us here is the character table of $\PSL(2,\Z_n)$. Let us begin with a concrete example, namely $n=3$. The character table of $\PSL(2,\Z_3)$ can be obtained by means of a computer algebra program like GAP \cite{GAP4} or MAGMA \cite{MR1484478}, with the result shown in Table~\ref{Tab:SL23}. Recall that each row of the character table corresponds to a different irreducible representation, while each column corresponds to a different conjugacy class. For a row labelled by representation $\rho_i$ and a column labelled by conjugacy class $[g]$, the corresponding entry in the table is $\chiG_i(g) = \mathrm{Tr}_{\rho_i}\, g$.\footnote{Note that here we are concerned with group characters $\chiG$, not to be confused with RCFT characters  $\chi$.} The first column, corresponding to the conjugacy class of the identity $[I]$, computes the dimension of  each irrep. We see that there are three one-dimensional irreps, three two-dimensional irreps, and a single three-dimensional irrep. This matches the $p=3$ row of Table \ref{Tab:SL2ZNRep}.

However, the character table clearly contains more information than this. Indeed, we may use the character table to reconstruct the full set of possible $T$ eigenvalues as follows. The group $\PSL(2,\Z_n)$ contains $S$ and $T$ elements, and in any given representation $\rho_i$ of $\PSL(2,\Z_n)$ we must have that $\rho_i(T^n) = 1$. The first step in obtaining the eigenvalues of $T$ in a representation $\rho_i$ is to identify the conjugacy classes of $T^m$ for $m = 0, 1, \dotsc, n-1$. Having done so, one then reads off the corresponding characters $\chiG_i(T^m)$ from the character table. Then with the characters in hand, one notes that the $\chiG_i(T^m)$ are related to the eigenvalues $\{t_a^{(i)}\}$ of $\rho_i(T)$  via a discrete Fourier transform,
\ie
\chiG_i(T^m)= \sum_{a=1}^d (t_a^{(i)})^m = \sum_{\ell=0}^{n-1} \nu^{(i)}_\ell \, e^{\frac{2\pi i \ell m}{n}} \,.
\fe
Here $\nu^{(i)}_m$ denotes the (possibly zero) degeneracy of $e^{\frac{2\pi i m}{k}}$ in the set of $\rho_i(T)$ eigenvalues (for a $d$-dimensional representation $\rho_i$, clearly $\sum_{\ell=0}^{n-1} \nu^{(i)}_\ell = d$). It is these $\nu^{(i)}_m$ which encode all of the data of the $\rho_i(T)$ eigenvalues. Performing the inverse discrete Fourier transform allows us to extract them,
\ie
\nu^{(i)}_m = \frac{1}{n} \sum_{p=0}^{n-1} \chiG_i(T^p) \, e^{-\frac{2\pi i m p}{n}} \, ,
\fe
and in this way we obtain the full set of $T$ eigenvalues---i.e. the set of allowed $\{h_j - \frac{c}{24}\}$.

Returning to the simple example of $n= 3$, let us now illustrate the procedure by identifying the exponents $\{h_j - \frac{c}{24}\}$ for the three-dimensional representation $\rho_7$. We first use GAP to find that $T$ is in the conjugacy class 3$b$, while $T^2$ is in the conjugacy class 3$a$. From the character table, we see that the corresponding characters are
\bea
\chiG_7(1) = 3~, \hspace{0.5 in}\chiG_7(T)=\chiG_7(T^2) = 0 ~.
\eea
Performing the inverse discrete Fourier transform then gives
\ie
\nu^{(7)}_0 = \nu^{(7)}_1 = \nu^{(7)}_2 = 1 \, ,
\fe
and hence the $\rho_i(T)$ eigenvalues are $\{ 1, \omega, \bar\omega \}$ for $\omega = e^{\frac{2 \pi i}{3}}$.  In other words, the exponents  $\{h_j - \frac{c}{24}\}$ are $\{ 0, \frac13, \frac23 \} \mod 1$.

\subsection{All allowed exponents and Wronskian indices}
\begin{table}[!tp]
\begin{center}
\makebox[\textwidth]{
\begin{tabular}{c|c}
$n$ &2d $\ell=0$ exponents
\\ \hline
6 & $\left\{{1 \over 3} , {5\over 6} \right\}$
\\
12 & $\left\{{1 \over 4} , {11\over 12} \right\}, \left\{{3 \over 4} , {5\over 12} \right\}$
\\
24 & $\left\{{11 \over 24} , {17\over 24} \right\},\left\{{5 \over 24} , {23\over 24} \right\}$
\\
60 & $\left\{{11 \over 60} , {59\over 60} \right\},\left\{{17 \over 60} , {53\over 60} \right\}, \left\{{23 \over 60} , {47\over 60} \right\}, \left\{{29 \over 60} , {41\over 60} \right\}$
\end{tabular}
}\vspace{10pt}
\makebox[\textwidth]{
\begin{tabular}{c|c}
$n$ & 2d $\ell=2 $ exponents
\\ \hline
6 & $\left\{{2 \over 3} , {1\over 6} \right\}$
\\
12 & $\left\{{1 \over 4} , {7\over 12} \right\}, \left\{{3 \over 4} , {1\over 12} \right\}$
\\
24 & $\left\{{1 \over 24} , {19\over 24} \right\},\left\{{7 \over 24} , {13\over 24} \right\}$
\\
60 & $\left\{{1 \over 60} , {49\over 60} \right\}, \left\{{7 \over 60} , {43\over 60} \right\},\left\{{19 \over 60} , {31\over 60} \right\}, \left\{{13 \over 60} , {37\over 60} \right\}$
\end{tabular}
}\vspace{10pt}
\makebox[\textwidth]{
\begin{tabular}{c|c}
$n$ & 2d $\ell=4$ exponents
\\ \hline
2 & $\left\{0 , {1\over 2} \right\}$
 \\
8 & $\left\{{1 \over 8} , {3\over 8} \right\},\left\{{5 \over 8} , {7\over 8} \right\}$
\\
12 & $\left\{{1 \over 12} , {5\over 12} \right\}, \left\{{7 \over 12} , {11\over 12} \right\}$
\\
20 & $\left\{{1 \over 20} , {9\over 20} \right\}, \left\{{3 \over 20} , {7\over 20} \right\},\left\{{11 \over 20} , {19\over 20} \right\}, \left\{{13 \over 20} , {17\over 20} \right\}$
\end{tabular}
}
\end{center}
\caption{Possible exponents mod 1 for two-character theories with $\ell=0,2, 4$ mod 6.}
\label{tab:2dl02}
\end{table}
The strategy for obtaining the full set of allowed exponents at fixed number of characters $d$ is now clear---we simply repeat the procedure of Section \ref{sec:numerators} for all $n$  in the lists $\mathfrak{den}(d)$ given in Section \ref{sec:denoms}.\footnote{In fact, it suffices to carry out the procedure on the least common multiple of the set of all $n$. } We will summarize the final results of this procedure, organizing our results by the values of the Wronskian index (\ref{WronskianDetT}). More precisely, since we have thus far only constrained the possible exponents mod 1, the values of the Wronskian index are defined only mod 6.

Starting with two-dimensional vvmfs, we find that only even values of the Wronskian index are possible. For both $\ell=0,2$,  there are nine classes of possible exponents mod 1, given in Table \ref{tab:2dl02}. For $\ell = 4$ there are again nine classes, also given in Table \ref{tab:2dl02}. Of course, from what we have said thus far it does not follow that all of these exponents are realized by legitimate two-character RCFTs. This question will be addressed in the following section for the $\ell=0,2$ cases, for which the exponents below completely fix the corresponding MDE.

For three-character vvmfs, the Wronskian index is found to always be a multiple of three. For each of $\ell=0,3$ mod 6 there are a total of 34 classes of allowed exponents, given in Table \ref{tab:3dl3} of Appendix \ref{app:extraexps}.

For four-character vvmfs, the Wronskian index is found to always be even. For each of $\ell = 0,2,4$ mod 6, there are  30 allowed exponents, given in Table \ref{tab:4dl2} of Appendix \ref{app:extraexps}.

Finally, for five-character vvmfs the Wronskian index can take any value. For each of choice of $\ell$ mod 6 there are only three allowed exponents mod 1, given in Table \ref{tab:5dl1} of Appendix \ref{app:extraexps}.

\subsubsection*{Wronskian index}

As an obvious byproduct of the above results, we have succeeded in rederiving the classic result that odd values of $\ell$ are disallowed for 2-character RCFTs \cite{Naculich:1988xv}. We have also obtained a significant generalization---indeed, we have seen that a Wronskian index $\ell$ can be realized for a particular $d$ only if there exists a representation of $\PSL(2,\Z_n)$, for some $n$ in $\mathfrak{den}(d)$, such that $\det T$ is given by \eqref{WronskianDetT}.  Above we have found that for small $d$, the allowed values of $\ell$ are given in the following table:
\begin{center}
\begin{tabular}{c|cccc}
$d$ & 2 & 3 & 4 & 5
\\\hline
$\ell$ & $2\Z$ & $3\Z$ & $2\Z$ & $\Z$
\end{tabular}
\end{center}
Conversely, all values of $\ell \text{ mod } 6$ in the above table are realized by actual vvmfs.
This is because we can always multiply a vvmf by a factor of $j(\tau)^{\frac{x}{3}} (1728 - j(\tau))^{\frac{y}{2}}$ for arbitrary $x, y \in \Z_{\ge 0}$ to increase the Wronskian by a factor of $(2x + 3y)d$ without introducing spurious poles.  Since, as will be shown in the next section,  monic $\ell = 0$ solutions exist for all $d = 2, 3, 4, 5$ the claim follows.

\section{Classification of solutions to rigid MDEs}
\label{sec:mainsec2}
\label{sec:computerscan}
In the previous section, we used the representation theory of $\PSL(2, \ZZ_n)$ to obtain all possible exponents for RCFTs with characters $d \leq 5$. Thus far, the only ingredient of RCFT which we have used is the integrality of the characters, which ensured that the character vector transformed in a vector of $\PSL(2, \ZZ_n)$. The results of the previous section then amount to a classification of exponents of quasicharacters.
Of course, not every integral vvmf can be interpreted as \textit{actual} characters for some RCFT. As discussed in Section \ref{Sec:Additional}, in addition to integrality the requirements of positivity and existence of a vacuum will be imposed.
To the best of our knowledge, there does not exist a purely representation-theoretic way of imposing these additional constraints. As such, our approach to the problem will instead be to explicitly solve the corresponding MDE order-by-order in $q$ for each choice of exponents,  and then to explicitly check for positivity and existence of a vacuum. For this to be a viable strategy, we should restrict ourselves to situations in which the exponents obtained above completely fix the MDE, i.e. to \textit{rigid} MDEs in the language of Section \ref{sec:MDE}. As reviewed there, monic degree $d \leq 5$ MDEs are all rigid, so in these cases the order-by-order check can be carried out. The case of 2d $\ell=2$ is also rigid, and can be addressed by similar means.

While solutions to monic MDEs of low order $d = 2, 3$ have been completely classified, a systematic study of $d \ge 4$ has been lacking.  This section gives the complete classification of potential $\ell=0$ theories with four and five characters.

\subsection{Two characters}

Beginning with the case of two characters, we find the results shown in Table \ref{table:ell=02} for $\ell=0, 2$.
These results match with the well-known results obtained in \cite{Mathur:1988na,Hampapura:2015cea}. The tables are color-coded as follows. Rows marked in green are theories which admit a positive integer Verlinde formula, i.e.  theories for which there are exactly two modules with distinct characters that transform under S- and T-matrices furnishing an MTC structure. On the other hand, rows marked in blue are theories with a negative integer Verlinde formula---these cases should be thought of as intermediate VOAs, including for example the $c={38 \over 5}$ entry, which corresponds to the $E_{7 \half }$ intermediate VOA. Finally, rows marked in red admits multiple vacuum interpretations; for example, the $(c,h)=({2 \over 5}, {1\over 5})$ theory can alternatively be interpreted as a theory with $(c,h)=(-{22 \over 5}, -{1\over 5})$, which we recognize as the Lee-Yang minimal model $\cM_{2,5}$. Note that for the choice of vacuum giving rise to $(c,h)=({2 \over 5}, {1\over 5})$, one has negative integer Verlinde formula. However we do not mark this case in blue, since the sign of Verlinde is a vacuum dependent statement. Indeed, for the rest of this paper we do not bother with the sign of Verlinde for characters with multiple vacuum interpretations, since it is vacuum dependent.

All of the theories in the $\ell=0$ table are known. We have already discussed the first and last entries above---the remaining seven entries are the $(A_1)_1$, $(A_2)_1$, $(G_2)_1$,  $(D_4)_1$, $(F_4)_1$, $(E_6)_1$, and $(E_7)_1$ WZW models, respectively. As for the $\ell=2$ entries, these are also well-known. The actual theories realizing these characters are cosets of meromorphic $c=24$ theories by the $\ell=0$ theories \cite{Gaberdiel:2016zke}.

\begin{table}[!tp]
\begin{minipage}{.5\linewidth}
\centering
\begin{tabular}{c|c|c}
 $c$ &$h$ & $n_J$
 \\\hline
 \rowcolor{red!20}
$ \frac{2}{5}$ & $\frac{1}{5}$  & 1  \\
\rowcolor{green!20}
 1 & $\frac{1}{4}$  & 3\\
 2 & $1 \over 3$ & 8 \\
 \rowcolor{green!20}
 $\frac{14}{5}$ & $\frac{2}{5}$ & 14 \\
 4 & $\half$  &  28 \\
 \rowcolor{green!20}
$\frac{26}{5}$ & $\frac{3}{5}$ & 52\\
 6 & $2 \over 3$  & 78\\
 \rowcolor{green!20}
 7 & $\frac{3}{4}$ &  133\\
  \rowcolor{blue!20}
$ \frac{38}{5}$ & $\frac{4}{5}$ & 190 \\
\end{tabular}
\end{minipage}
 \begin{minipage}{.5\linewidth}
\centering
\begin{tabular}{c|c|c}
 $c$ &$h$ & $n_J$
 \\\hline
   \rowcolor{blue!20}
${82 \over 5}$ & $6 \over 5$ & 410  \\
 \rowcolor{green!20}
17 & $5 \over 4$ &  323\\
 18 & $4 \over 3$  & 234 \\
  \rowcolor{green!20}
$94 \over 5$ & $7 \over 5$  & 188\\
 20 & $3 \over 2$ & 140 \\
  \rowcolor{green!20}
$106 \over 5$ & $8 \over 5$ & 106 \\
$22$& $ 5\over 3$&  88 \\
 \rowcolor{green!20}
 23 & $7\over 4$  & 69 \\
   \rowcolor{blue!20}
$118 \over 5$ & $9 \over 5$  & 59 \\
\end{tabular}
 \end{minipage}
 \caption{Potential $(d,\ell) =(2,0)$ (left) and $(d,\ell) =(2,2)$ (right) theories. Rows colored red denote theories which admit multiple vacuum interpretations. Here we have chosen a single possible interpretation, and $n_J$ gives the number of spin-1 currents in that interpretation. Rows colored green (blue) denote cases with a positive (negative) integer Verlinde formula. }
  \label{table:ell=02}
\end{table}

\subsection{Three characters}

 \begin{table}[!htp]
\makebox[\textwidth]{
\begin{minipage}[t]{.3\linewidth}
\centering
\begin{tabular}[t]{c|c|c}
 $c$ & $h_i$& $n_J$
\\\hline
 $4$ & ${1 \over 3}$ , $2 \over 3$ &16
 \\
  $4$ & ${2 \over5}$ , $3 \over 5$&24
  \\
 ${12}$ & ${1 \over 3}$ , $5 \over 3$& 318
 \\
 ${12}$ & ${2 \over 3}$ , $4 \over 3$& 156
 \\
  ${12}$ & ${3 \over 5}$ , $7 \over 5$&222
    \\
      \rowcolor{green!20}
 ${13}$ & ${5 \over 8}$ , ${3 \over 2}$& 273
  \\
    \rowcolor{green!20}
 ${14}$ & ${3 \over 4}$ , ${3 \over 2}$& 266
  \\
    \rowcolor{green!20}
 ${15}$ & ${7 \over 8}$ , ${3 \over 2}$& 255
  \\
   \rowcolor{green!20}
 ${17}$ & ${9 \over 8}$ , ${3 \over 2}$& 221
  \\
  \rowcolor{green!20}
 ${18}$ & ${5 \over 4}$ , ${3 \over 2}$& 198
   \\
     \rowcolor{green!20}
 ${19}$ & ${11 \over 8}$ , ${3 \over 2}$& 171
  \\
 ${20}$ & ${1 \over 3}$ , $8 \over 3$&728
   \\
 ${20}$ & ${2 \over 3}$ , $7\over 3$&890
   \\
 ${20}$ & ${4 \over 3}$ , $5\over 3$&80
  \\
 ${20}$ & ${7 \over 5}$ , $8 \over 5$& 120
 \\
   \rowcolor{green!20}
 ${21}$ & ${13 \over 8}$ , ${3 \over 2}$&105
  \\
    \rowcolor{green!20}
 ${22}$ & ${7 \over 4}$, ${3 \over 2}$& 66
 \\
   \rowcolor{green!20}
    ${22}$ & ${3 \over 4}$ , ${5 \over 2}$& 1298
  \\
    \rowcolor{green!20}
 ${23}$ & ${15 \over 8}$ , ${3 \over 2}$& 23
 \\
   \rowcolor{green!20}
 ${23}$ & ${7 \over 8}$ , ${5 \over 2}$& 2323
  \\
 ${28}$ & ${2 \over 3}$ , $10\over 3$&1948
  \\
 ${36}$ & ${2 \over 3}$ , $13\over 3$& 3384
\end{tabular}\end{minipage}
\begin{minipage}[t]{.3\linewidth}
\centering
\begin{tabular}[t]{c|c|c}
 $c$ & $h_i$& $n_J$
\\\hline
      \rowcolor{green!20}
 ${25 \over 2}$ & ${9\over 16}$ , ${3 \over 2}$& 275
   \\
     \rowcolor{green!20}
 ${27 \over 2}$ & ${11\over 16}$ , ${3 \over 2}$& 270
   \\
     \rowcolor{green!20}
  ${29 \over 2}$ & ${13 \over 16}$ , $3 \over 2$& 261
 \\
 \rowcolor{green!20}
 ${31 \over 2}$ & ${15 \over 16}$ , $3 \over 2$& 248
  \\
    \rowcolor{green!20}
 ${33 \over 2}$ & ${17\over 16}$ , ${3 \over 2}$& 231
  \\
    \rowcolor{green!20}
 ${35 \over 2}$ & ${19\over 16}$ , ${3 \over 2}$& 210
 \\
   \rowcolor{green!20}
${37 \over 2}$ & ${21 \over 16}$ , $3 \over 2$& 185
\\
  \rowcolor{green!20}
${39 \over 2}$ & ${23 \over 16}$ , $3 \over 2$& 156
 \\
   \rowcolor{green!20}
 ${41 \over 2}$ & ${25\over 16}$ , ${3 \over 2}$&123
  \\
    \rowcolor{green!20}
 ${43 \over 2}$ & ${27\over 16}$ , ${3 \over 2}$& 86
 \\
   \rowcolor{green!20}
${45 \over 2}$ & $29 \over 16$ , ${3 \over 2}$& 45
\\
  \rowcolor{green!20}
 ${47 \over 2}$ & $31 \over 16$ , ${3 \over 2}$&0
\end{tabular}\end{minipage}
\begin{minipage}[t]{.3\linewidth}
\centering
\begin{tabular}[t]{c|c|c}
 $c$ & $h_i$&$n_J$
\\\hline
  \rowcolor{red!20}
 ${4 \over 5}$ & ${1 \over 5}$ , $2 \over 5$ & 2
 \\
 ${12 \over 5}$ & ${1 \over 5}$ , $3 \over 5$  &3
 \\
  ${28 \over 5}$ & ${2 \over 5}$ , $4 \over 5$&28
\\
 ${36 \over 5}$ & ${3 \over5}$ , $4 \over 5$& 144
\\
  ${44 \over 5}$ & ${2 \over 5}$ , $6 \over 5$&220
  \\
  ${52 \over 5}$ & ${3 \over 5}$ , $6 \over 5$&104
 \\
  ${68 \over 5}$ & ${4 \over 5}$ , $7 \over 5$& 136
\\
  ${76 \over 5}$ & ${3 \over 5}$ , $9 \over 5$&437
\\
 ${76 \over 5}$ & ${4 \over 5}$ , $8 \over 5$&380
\\
${84 \over 5}$ & ${6 \over 5}$ , $7 \over 5$&336
 \\
  ${92 \over 5}$ & ${6 \over 5}$ , $8 \over 5$& 92
\\
 ${108 \over 5}$ & ${4 \over 5}$ , $12 \over 5$& 1404
\\
 ${108 \over 5}$ & ${7 \over 5}$ , $9 \over 5$ & 27
\\
 ${116 \over 5}$ & ${8 \over 5}$ , $9 \over 5$ &58
\\
 ${164 \over 5}$ & ${11\over 5}$ , $12 \over 5$&0
\end{tabular}\end{minipage}
   \begin{minipage}[t]{.3\linewidth}
\centering
\begin{tabular}[t]{c|c|c}
 $c$ & $h_i$& $n_J$
\\\hline
  \rowcolor{red!20} ${4 \over 7}$ &   ${1 \over 7}$, $3 \over 7$  &1
\\
\rowcolor{blue!20}
 ${12 \over 7}$ & ${2 \over 7}$ , $3 \over 7$& 6
\\
\rowcolor{blue!20}
  ${44 \over 7}$ & ${4 \over 7}$ , $5 \over 7$& 88
\\
\rowcolor{blue!20}
  ${52 \over 7}$ & ${4 \over 7}$ , $6 \over 7$& 156
\\
\rowcolor{blue!20}
  ${60 \over 7}$ & ${3 \over 7}$ , $8 \over 7$& 210
\\
\rowcolor{blue!20}
 ${68 \over 7}$ & ${3 \over 7}$ , $9 \over 7$& 221
\\
\rowcolor{blue!20}
  ${100 \over 7}$ & ${4 \over 7}$ , $12 \over 7$& 380
\\
\rowcolor{blue!20}
 ${100 \over 7}$ & ${5 \over 7}$ , $11 \over 7$& 325
\\
\rowcolor{blue!20}
 ${108 \over 7}$ & ${6 \over 7}$ , $11 \over 7$& 378
\\
\rowcolor{blue!20}
 ${108 \over 7}$ & ${4 \over 7}$ , $13 \over 7$&456
\\
\rowcolor{blue!20}
 ${116 \over 7}$ & ${8 \over 7}$ , $10 \over 7$&348
\\
\rowcolor{blue!20}
 ${124 \over 7}$ & ${9 \over 7}$ , $10 \over 7$& 248
\\
\rowcolor{blue!20}
 ${156 \over 7}$ & ${11 \over 7}$ , $12 \over 7$& 78
\\
\rowcolor{blue!20}
 ${156 \over 7}$ & ${5 \over 7}$ , $18 \over 7$ &1248
\\
\rowcolor{blue!20}
 ${164 \over 7}$ & ${11 \over 7}$ , $13 \over 7$ & 41
\\
\rowcolor{blue!20}
 ${236 \over 7}$ & ${16 \over 7}$ , $17 \over 7$ & 0
\end{tabular}
  \end{minipage} }
  \caption{
  List of all potential  $(d,\ell) = (3, 0)$ theories, not including the infinite series of theories $\mathrm{Spin}(n)_1$. Rows colored red denote theories which admit multiple vacuum interpretations. Here we have chosen a single possible interpretation, and $n_J$ gives the number of spin-1 currents in that interpretation. Rows colored green (blue) denote cases with a positive (negative) integer Verlinde formula. }
    \label{tab:3dl0tabimprim1}
\end{table}

We now proceed to the case of three characters. In this case we observe something which is unlike any of the other cases studied in this paper: namely, an infinite set of legitimate theories. This set of theories decomposes into the infinite series $\mathrm{Spin}(n)_1$ for $n \neq 0$ mod 8, with central charge and chiral dimensions given by\footnote{Note that we do not obtain the cases of  ${\rm Spin}(8n)_1$, since those cases are reducible. For example, the ${\rm Spin}(8)_1$ case is just the $(D_4)_1$ WZW model, which is really a two-character theory.}
\bea
\mathrm{Spin}(n)_1: \hspace{0.6 in} (c,h_1,h_2) = \left({n\over2},\, \half, {n \over 16} \right)~,
\eea
and a finite number of remaining theories. The remaining theories are given in Table \ref{tab:3dl0tabimprim1}; for a closely related table, see \cite{franc2020classification}.  Some of these theories are familiar, including the  $(A_4)_1$, $(E_8)_2$, $(G_2)_1^{\otimes2}$, $(F_4)_1^{\otimes2}$, and $(E_{7 \half})_1^{\otimes2}$ theories with respective data $(c,h_1,h_2) = (4, {2\over 5},{3\over 5}),({31 \over 2}, {15\over 16},{3\over 2}),({28 \over 5}, {2\over 5},{4\over 5})$, $({52 \over 5}, {3\over 5},{6\over 5})$, and $({76 \over 5}, {4\over 5},{8\over 5})$.

In Table \ref{tab:3dl0tabimprim1} we observe two entries which allow for multiple vacuum interpretations. In both cases, there exists an interpretation in which the theory has no spin-1 currents. In particular, the $c=({4\over 7},{1\over 7},{3 \over 7})$ entry admits an alternative interpretation as a $(c, h_1, h_2) = (-{68 \over 7}, -{2 \over 7},-{3 \over 7})$ theory, which is the non-unitary minimal model $\cM_{2,7}$. Likewise the $(c, h_1, h_2)=({4\over 5},{1\over 5},{2 \over 5})$ entry admits an interpretation as a $(c, h_1, h_2) =(-{44 \over 5},-{2 \over 5}, - {1 \over 5})$ theory, which is $(\cM_{2,5})^{\otimes 2}$.

Including theories with a single vacuum interpretation, we find a total of 6 theories without Kac-Moody. Five of them are in Table \ref{tab:3dl0tabimprim1}, and one of them is the Ising model, i.e. $\mathrm{Spin}(1)_1$. We have already given an interpretation to three of these theories. The theory with $(c, h_1, h_2)=({47 \over 2}, {31 \over 16}, {3 \over 2})$ can also be given an interpretation, namely as the Baby Monster CFT. Just as the Ising and Baby Monster characters are ``dual" in the sense that the inner product of their characters gives the $j$-function, so too are the $c={236 \over 7}, {164 \over 5}$ theories dual to $\cM_{2,7}$ and $(\cM_{2,5})^{\otimes 2}$. This spectrum of six theories without Kac-Moody was previously identified in \cite{Hampapura:2016mmz}.

In total, we observe 23 cases with positive integral Verlinde formula.

\subsection{Four characters}

We now proceed to the case of four characters, where the results are entirely new. The list of possible four-character $\ell=0$ theories satisfying positivity and existence of a vacuum is given in Table~\ref{Tab:Unitary}.
Many elements of the list can be recognized as familiar theories---this includes the
$(A_1)_3$, $(A_2)_2$, $(A_2)_6$, $(C_3)_1$, $(A_5)_1$, $(G_2)_2$, and $(F_4)_6$ WZW models, which correspond respectively to the theories with central charges $c={9 \over 5}, {16 \over 5}, {16 \over 3}, {21 \over 5}, 5, {14 \over 3}$ and ${104 \over 5}$. We also identify some product theories, including the $(A_1)_1 \times (C_5)_3$ WZW model with $(c,n_J )=({58 \over 3},58)$.
\begin{table}[!t]
\centering
\makebox[\textwidth]{
$\begin{array}[t]{c|c|c}
c & h_i & n_J
\\\hline
 1 & \frac{1}{12},\frac{1}{3},\frac{3}{4} & 1 \\
 2 & \frac{1}{6},\frac{1}{2},\frac{2}{3} & 2 \\
 3 & \frac{1}{4},\frac{1}{2},\frac{3}{4} & 9 \\
 5 & \frac{5}{12},\frac{2}{3},\frac{3}{4} & 35 \\
 6 & \frac{3}{7},\frac{5}{7},\frac{6}{7} & 48 \\
 18 & \frac{8}{7},\frac{9}{7},\frac{11}{7} & 144 \\
 19 & \frac{5}{4},\frac{4}{3},\frac{19}{12} & 133 \\
 21 & \frac{3}{4},\frac{3}{2},\frac{9}{4} & 399 \\
 21 & \frac{5}{4},\frac{3}{2},\frac{7}{4} & 63 \\
 22 & \frac{4}{3},\frac{3}{2},\frac{11}{6} & 22 \\
 23 & \frac{11}{12},\frac{5}{3},\frac{9}{4} & 575 \\
 23 & \frac{5}{4},\frac{5}{3},\frac{23}{12} & 23 \\
\end{array}$
\qquad
$
\begin{array}[t]{c|c|c}
c & h_i & n_J
\\\hline
\rowcolor{red!20}
{{2\over 3}} & {1\over 9} , {1 \over 3}, {2 \over 3} & 1
\\
\rowcolor{red!20}{{4\over 3}} & {2\over 9} , {1 \over 3}, {2 \over 3} & 4
\\
{{8\over 3}} & {1\over 3} , {4 \over 9}, {2 \over 3} & 12
\\
\rowcolor{blue!20}
{{10\over 3}} & {1\over 3} , {5 \over 9}, {2 \over 3} & 15
\\
 \rowcolor{green!20}{{14\over 3}} & {1\over 3} , {2 \over 3}, {7 \over 9} & 14
\\
{{16\over 3}} & {1\over 3} , {2 \over 3}, {8 \over 9} & 8
\\
{{56\over 3}} & {10\over 9} , {4 \over 3}, {5 \over 3} & 28
\\
{{56\over 3}} & {2\over 3} , {4 \over 3}, {19 \over 9} & 420
\\
 \rowcolor{green!20}{{58\over 3}} & {11\over 9} , {4 \over 3}, {5 \over 3} & 58
\\
 \rowcolor{green!20}{{58\over 3}} & {2\over 3} , {4 \over 3}, {20 \over 9} & 638
\\
\rowcolor{blue!20}
{{62\over 3}} & {4\over 3} , {13 \over 9}, {5 \over 3} & 93
\\
{{64\over 3}} & {4\over 3} , {14 \over 9}, {5 \over 3} & 96
\\
{{68\over 3}} & {2\over 3} , {4 \over 3}, {25 \over 9} & 2278
\\
{{68\over 3}} & {4\over 3} , {5 \over 3}, {16 \over 9} & 68
\\
\rowcolor{blue!20}
{{70\over 3}} & {2\over 3} , {4 \over 3},{26 \over 9} & 2730
\\
\rowcolor{blue!20}
{{70\over 3}} & {4\over 3} , {5 \over 3}, {17 \over 9} & 35
\end{array}
$
\qquad
$\begin{array}[t]{c|c|c}
c & h_i & n_J
\\\hline
\rowcolor{red!20}
\frac{3}{5} & \frac{1}{20},\frac{1}{4},\frac{4}{5} &1 \\
\rowcolor{red!20}
  \frac{4}{5} & \frac{1}{15},\frac{2}{5},\frac{2}{3} & 0 \\
 \frac{6}{5} & \frac{1}{10},\frac{1}{2},\frac{3}{5} & 0 \\
 \rowcolor{red!20}
  \frac{6}{5} & \frac{1}{5},\frac{2}{5},\frac{3}{5} & 3 \\
\frac{8}{5} & \frac{2}{15},\frac{1}{3},\frac{4}{5} & 4 \\
\rowcolor{green!20}\frac{9}{5}~ & \frac{3}{20},\frac{2}{5},\frac{3}{4} & 3 \\
\rowcolor{blue!20}
 \frac{12}{5} & \frac{1}{5},\frac{2}{5},\frac{4}{5} & 8 \\
 \frac{16}{5} & \frac{4}{15},\frac{3}{5},\frac{2}{3} & 8 \\
 \frac{18}{5} & \frac{3}{10},\frac{1}{2},\frac{4}{5} & 18 \\
\rowcolor{green!20}\frac{21}{5}~ & \frac{7}{20},\frac{3}{5},\frac{3}{4} & 21 \\
\rowcolor{blue!20}
 \frac{24}{5} & \frac{2}{5},\frac{3}{5},\frac{4}{5} & 36 \\
 \frac{28}{5} & \frac{7}{15},\frac{2}{3},\frac{4}{5} & 56 \\
 \rowcolor{blue!20}
 \frac{33}{5} & \frac{11}{20},\frac{3}{4},\frac{4}{5} & 99 \\
 \frac{42}{5} & \frac{2}{5},\frac{4}{5},\frac{6}{5} & 42 \\
  \frac{78}{5} & \frac{3}{5},\frac{6}{5},\frac{9}{5} & 156 \\
 \frac{78}{5} & \frac{4}{5},\frac{6}{5},\frac{8}{5} & 78 \\
 \rowcolor{blue!20}
 \frac{87}{5} & \frac{6}{5},\frac{5}{4},\frac{29}{20} & 261 \\
 \frac{92}{5} & \frac{6}{5},\frac{4}{3},\frac{23}{15} & 184 \\
\end{array}$
\qquad
$\begin{array}[t]{c|c|c}
c & h_i & n_J
\\\hline
\rowcolor{blue!20}
 \frac{96}{5} & \frac{3}{5},\frac{7}{5},\frac{11}{5} & 276 \\
 \rowcolor{blue!20}
 \frac{96}{5} & \frac{6}{5},\frac{7}{5},\frac{8}{5} & 144 \\
 \rowcolor{green!20}\frac{99}{5}~ & \frac{5}{4},\frac{7}{5},\frac{33}{20} & 99 \\
 \frac{102}{5} & \frac{6}{5},\frac{3}{2},\frac{17}{10} & 102 \\
 \frac{104}{5} & \frac{4}{3},\frac{7}{5},\frac{26}{15} & 52 \\
 \rowcolor{blue!20}
 \frac{108}{5} & \frac{4}{5},\frac{8}{5},\frac{11}{5} & 204 \\
 \rowcolor{blue!20}
 \frac{108}{5} & \frac{6}{5},\frac{8}{5},\frac{9}{5} & 72 \\
\rowcolor{green!20}
\frac{111}{5} & \frac{5}{4},\frac{8}{5},\frac{37}{20} & 37 \\
 \frac{112}{5} & \frac{13}{15},\frac{5}{3},\frac{11}{5} & 210 \\
 \frac{112}{5} & \frac{6}{5},\frac{5}{3},\frac{28}{15} & 56 \\
 \frac{114}{5} & \frac{4}{5},\frac{8}{5},\frac{12}{5} & 570 \\
 \frac{114}{5} & \frac{9}{10},\frac{3}{2},\frac{12}{5} & 1938 \\
 \frac{114}{5} & \frac{7}{5},\frac{3}{2},\frac{19}{10} & 0 \\
 \frac{114}{5} & \frac{7}{5},\frac{8}{5},\frac{9}{5} & 57 \\
 \frac{116}{5} & \frac{14}{15},\frac{8}{5},\frac{7}{3} & 1566 \\
 \frac{116}{5} & \frac{4}{3},\frac{8}{5},\frac{29}{15} & 0 \\
 \rowcolor{blue!20}
 \frac{117}{5} & \frac{19}{20},\frac{7}{4},\frac{11}{5} & 325 \\
 \rowcolor{blue!20}
 \frac{117}{5} & \frac{6}{5},\frac{7}{4},\frac{39}{20} & 39 \\
 \rowcolor{blue!20}
 \frac{123}{5} & \frac{5}{4},\frac{9}{5},\frac{41}{20} & 0 \\
\end{array}$
\qquad
$\begin{array}[t]{c|c|c}
c & h_i & n_J
\\\hline
\rowcolor{red!20} \frac{6}{7} & \frac{1}{7},\frac{2}{7},\frac{5}{7} & 2 \\
 \frac{18}{7} & \frac{1}{7},\frac{3}{7},\frac{6}{7} & 3 \\
 \frac{30}{7} & \frac{3}{7},\frac{4}{7},\frac{5}{7} & 30 \\
 \frac{138}{7} & \frac{9}{7},\frac{10}{7},\frac{11}{7} & 138 \\
 \frac{150}{7} & \frac{6}{7},\frac{11}{7},\frac{15}{7} & 300 \\
 \frac{150}{7} & \frac{8}{7},\frac{11}{7},\frac{13}{7} & 25 \\
 \frac{162}{7} & \frac{9}{7},\frac{12}{7},\frac{13}{7} & 54 \\
\end{array}$
}
\caption{List of all potential  $(d,\ell)=(4,0)$ theories. Rows colored red denote theories which admit multiple vacuum interpretations. Here we have chosen a single possible interpretation, and $n_J$ gives the number of spin-1 currents in that interpretation. Rows colored green (blue) denote cases with a positive (negative) integer Verlinde formula. }
\label{Tab:Unitary}
\end{table}

We find six cases which admit multiple vacuum interpretations, given in red, and in all cases there is a choice of vacuum with no Kac-Moody. In particular, the $c={3 \over 5}, {4 \over 5}, {6 \over 5}, {6\over 7}, {2\over 3},$ and ${4 \over 3}$ entries could be interpreted as $c=-{3 \over 5}, {4 \over 5}, -{66 \over 5}, -{144\over 7}, -{46\over 3},$ and $-{44 \over 3}$ theories with no Kac-Moody. In fact, the $c = -{3 \over 5}$ and $-{46\over 3}$ theories obtained in this way are respectively the non-unitary $\cM_{3,5}$ and $\cM_{2,9}$ minimal models, while $c = {4 \over 5}$ is the three-state Potts model. Including cases with only a single vacuum interpretation, we see that there are in total ten possible cases with no Kac-Moody. Among them, the $(c,n_J)=({116 \over 5},0)$ case matches with the theory denoted $VF_{24}^\natural$ in \cite{Fischerpaper,Bae:2020pvv}. This theory is associated with the $3.Fi_{24}'$ subgroup of the monster group, and the inner product of the three-state Potts and $VF_{24}^\natural$ characters gives the Monster CFT.

In two cases, namely those with $c = \frac{8}{5}$ and $\frac{6}{7}$, the vacuum characters reveal the existence of $n_J = 4$ and $n_J = 2$ spin-one conserved currents, for which the minimal Sugawara central charge is 2, exceeding both $c = \frac{8}{5}$ and $c = \frac{6}{7}$ \cite{Benjamin:2020zbs}. For $c = \frac{6}{7}$ this issue can be circumvented by changing the vacuum interpretation, which gives either a theory with $c=-{18\over 7}$ and $n_J=1$ or the $c=-{114\over 7}$ theory with no Kac-Moody mentioned above. However, in the case of $c = \frac{8}{5}$ there is no alternative vacuum interpretation. This case should be thought of as a non-unitary theory arising from an intermediate VOA, and can be identified with the theory discussed in e.g. Section  2.4. of \cite{Harvey:2019qzs}.

Among the theories identified, seven have a Verlinde formula that gives non-negative integral fusion coefficients, given in green in Table~\ref{Tab:Unitary}.

\subsection{Five characters}

\begin{table}[!t]
\centering
\begin{minipage}[t]{.4\linewidth}
\scalebox{1}{
$
\begin{array}[t]{c|c|c}
c & h_i & n_J
\\\hline
\rowcolor{red!20}
{8 \over 5} & {1\over 5} , {2 \over 5} , {3\over 5}, {4\over 5} & 4
\\
{32 \over 5} & {1\over 5} , {3 \over 5} , {4\over 5}, {7\over 5} & 82
\\
{32 \over 5} & {2\over 5} , {3 \over 5} , {4\over 5}, {6\over 5} & 80
\\
{56 \over 5} & {2\over 5} , {4 \over 5} , {6\over 5}, {8\over 5} & 56
\\
{56 \over 5} & {3\over 5} , {4 \over 5} , {6\over 5}, {7\over 5} & 28
\\
{104 \over 5} & {3\over 5} , {6 \over 5} , {9\over 5}, {12\over 5} & 208
\\
{104 \over 5} & {4\over 5} , {7 \over 5} , {8\over 5}, {11\over 5} & 520
\\
{104 \over 5} & {6\over 5} , {7 \over 5} , {8\over 5}, {9\over 5} & 0
\\
{128 \over 5} & {4\over 5} , {7 \over 5} , {11\over 5}, {13\over 5} & 28
\\
{152 \over 5} & {4\over 5} ,{8 \over 5},  {12\over 5}, {16 \over 5} & 760
\end{array}
$
}
\end{minipage}
\begin{minipage}[t]{.4\linewidth}
\centering
\scalebox{1}{
$
\begin{array}[t]{c|c|c}
c & h_i & n_J
\\\hline
\rowcolor{red!20}
{8 \over 11} & {1\over 11} , {3 \over 11} , {6\over 11}, {10 \over 11} & 1
\\
\rowcolor{blue!20}
{32 \over 11} & {1\over 11} , {4 \over 11} , {7\over 11}, {13 \over 11} & 10
\\
\rowcolor{blue!20}
{32 \over 11} & {2\over 11} , {4 \over 11} , {7\over 11}, {12 \over 11} & 8
\\
\rowcolor{blue!20}
{56 \over 11} & {4\over 11} , {7 \over 11} , {9\over 11}, {10 \over 11} & 28
\\
\rowcolor{blue!20}
{80 \over 11} & {5\over 11} , {8 \over 11} , {10\over 11}, {12 \over 11} & 120
\\
\rowcolor{green!20}
{104 \over 11} & {6\over 11} , {9 \over 11} ,{12\over 11}, {13 \over 11} & 52
\\
\rowcolor{blue!20}
{128 \over 11} & {5\over 11} , {8 \over 11}  {15\over 11},,{17 \over 11} & 248
\\
\rowcolor{blue!20}
{128 \over 11} & {6\over 11} , {8 \over 11} , {15\over 11}, {16 \over 11} & 224
\\
\rowcolor{blue!20}
{224 \over 11} & {5\over 11} , {14 \over 11} , {18\over 11}, {28 \over 11} & 528
\\
\rowcolor{blue!20}
{224 \over 11} & {14\over 11} , {16 \over 11} , {17\over 11}, {18 \over 11} & 112
\\
\rowcolor{green!20}
{248 \over 11} & {10\over 11} , {16 \over 11} , {20\over 11}, {24 \over 11} & 248
\\
\rowcolor{green!20}
{248 \over 11} & {13\over 11} , {16 \over 11} , {20\over 11},{21 \over 11} & 0
\\
\rowcolor{blue!20}
{344 \over 11} & {10\over 11} , {19 \over 11} , {27\over 11}, {34 \over 11} & 946
\end{array}
$
}
\end{minipage}
\caption{List of all potential  $(d,\ell)=(5, 0)$ theories. Rows colored red denote theories which admit multiple vacuum interpretations. Here we have chosen a single possible interpretation, and $n_J$ gives the number of spin-1 currents in that interpretation. Rows colored green (blue) denote cases with a positive (negative) integer Verlinde formula. }
\label{tab:5dl0physical}
\end{table}

Finally we proceed to the case of five characters. The list of possible five-character $\ell=0$ theories satisfying positivity and existence of a vacuum are given in Table~\ref{Tab:Unitary}. Three theories on this list admit a positive integral Verlinde formula, which include the $(F_4)_2$ and $(E_8)_3$ WZW models with respective central charges $c={104 \over 11}$ and $c = {248 \over 11}$. We also recognize the $(c, n_J) = ({56 \over 5}, 28)$ theory as $(D_4)_4$.

We observe two cases which admit multiple vacuum interpretations, and in both cases there is a choice in which there is no Kac-Moody. Indeed, the $c={8 \over 5}$ and ${8 \over 11}$ cases can be respectively interpreted as $ c= -{88 \over 5}$ and $-{232 \over 11}$ theories with no Kac-Moody. The latter in particular is the $\cM_{2,11}$ non-unitary minimal model. Including theories with only a single vacuum interpretation, we find a total of four potential theories without Kac-Moody.

\section{The Mathur-Sen approach}
\label{sec:MathurSen}
Thus far we have made crucial use of the fact that integral characters transform in representations of finite subgroups of $\PSL(2, \ZZ_n)$. However, there is an alternative approach, initiated by Mathur and Sen in \cite{mathur1989group}, based on the observation that known RCFT characters arise from MDEs with \textit{finite monodromy group}.

To understand this, let us begin by introducing the notion of a monodromy group associated to an MDE. The class of rigid MDEs that we consider only have singularities at the points $\tau = i\infty,i,$ and $\omega:=e^{2\pi i/3}$ in the fundamental domain. Upon the change of variables $x=j(\tau)/1728$, these get mapped to $x=\infty,1,0$. The domain $D := \mathbb{P}^1/\{1,0,\infty\}$ of $x$ has a fundamental group generated by three loops $\g_i$ surrounding the singular points,
\bea
\pi_1(D) = \langle \g_0, \g_1, \g_\infty \,| \, \,\g_0 \g_1 \g_\infty =1 \rangle~.
\eea
For instance, $\gamma_\infty$ surrounding the point at infinity can be taken as the small loop $\gamma_\infty: x^{-1}(s) = \varepsilon e^{2 \pi i s}$, $s \in [0,1]$.
We introduce a $d$-dimensional linear representation of the fundamental group
\bea
M:\,\, \pi_1(D)\rightarrow \GL(d,\CC)\,,
\eea
where the matrices $M_i := M(\g_i)$ can be obtained by considering the transformation of the solutions of the MDE when traversing each loop. For instance, let us choose basis of solutions of the MDE as a series expansion around $x=\infty$\footnote{Using the familiar Fourier expansion of $j(\tau) = 1728 x$,
\bea
j(\tau) = \frac1q + 744 + 196884 q + \cdots
\eea
the series solution \eqref{eq:seriessol} yields the familiar Fourier series in Eq.~\eqref{eq:fseriessol}.}
\bea
\chi_i(x) = x^{-\alpha_i} ( c_{i,0} +  c_{i,1}\, x^{-1} + c_{i,2}\, x^{-2} + \cdots)\,,
\label{eq:seriessol}
\eea
Upon traversing $\gamma_\infty$ we have
\bea
\begin{pmatrix} \chi_1(x) \\ \vdots \\ \chi_d(x) \end{pmatrix} \longrightarrow M_\infty \begin{pmatrix} \chi_1(x) \\ \vdots \\ \chi_d(x) \end{pmatrix}\,,
\eea
 with $ M_\infty = \text{diag} (e^{2\pi i \alpha_1}, \cdots, e^{2\pi i \alpha_d})$. $ M_0$ and $M_1$ are  similarly defined, although generically they will not be diagonal in this basis.
The monodromy group $\cM$ of the MDE is then defined to be the group  of complex $d\times d$ matrices generated by
\bea
\label{eq:mongroupdef}
\cM = \langle M_0, M_1, M_\infty \in \GL(d,\CC) \,| \, \,M_0 M_1 M_\infty =1 \rangle ~.
\eea

Using the inverse change of variables
one can show that traversing the loop $\gamma_\infty$ sends $\tau \rightarrow \tau + 1$, and $\gamma_1$ takes $\tau \to -1/\tau$.
On the upper half plane, $\gamma_1$ and $\gamma_0$ traverse a half circle and a third of a circle, respectively.
Because characters are holomorphic and hence single-valued at $\tau = i$ and $\omega$, traversing a full circle is trivial, so  $M_1^2=1$ and $M_0^3 = 1$.
The monodromy matrices of the MDE may then be identified with the modular S- and T-transformations of the characters as follows,
\bea
M_\infty = T~,  \hspace{0.5 in}  M_1 = S~, \hspace{0.5 in}  M_0 = (ST)^{-1}~.
\eea
In general, these S- and T-matrices do not represent $\PSL(2,\ZZ)$ faithfully. For instance, due to the rationality of the exponents $\alpha_i$, $T$ has finite order $n$ (i.e. $T^n = 1$).
So the monodromy trivializes the action of the congruence subgroup $\Gamma(n)$.
This provides the connection to the finite groups $\PSL(2,\ZZ_n) = \PSL(2,\ZZ)/\Gamma(n)$ studied in the previous section.

The monodromy group of a $d$-character RCFT must be a finite quotient of $\PSL(2,\Z)$ with a faithful $d$-dimensional representation, which in turn yields a representation of a finite subgroup of $\GL(d,\CC)$.  By studying representations of finite subgroups of $\GL(2,\CC)$ and $\GL(3,\CC)$, Mathur and Sen found that, apart from an infinite family, there are only a finite number of possibilities for theories with $d=2$ and $d=3$. From these possibilities they constrained the possible common denominators of $\{h_i - \frac{c}{24}\}$, since a common denominator $n$ implies that there exists an element of order $n$ in the monodromy group.  For instance, they proved that for $d=3$, the central charge must be $c \in \frac{\Z}{70}$. The result was a complete classification of $d=2$ theories and a partial classification of $d=3$ theories.

In this section we complete the $d=3$ analysis, as well as carry out the $d=4$ analysis. The results match precisely with those obtained in the previous sections, yielding a convincing check. Note that the finite subgroups of $\GL(d,\CC)$ are more complicated than $\PSL(2, \ZZ_n)$, which makes this approach less manageable at higher $d$. We catalogue the finite subgroups of $\SL(3,\CC)$ and $\SL(4,\CC)$ in Appendices \ref{app:SL3subs} and \ref{app:SL4subs}, from which the finite subgroups of  $\GL(3,\CC)$ and  $\GL(4,\CC)$ can be obtained by appropriate extension.

\subsection{Three characters}

We begin with the analysis of the three-character case. A key feature which we will utilize is the fact that all three-character $\ell=0$ MDEs can be recast into hypergeometric form under the variable transformation $x = \frac{j(\tau)}{1728}$ \cite{Naculich:1988xv,Mathur:1988gt,franc2016hypergeometric}. The monodromy of the generalized hypergeometric functions ${}_nF_{n-1}$ is a classic subject studied by many famous mathematicians including Schwarz, Klein, Gordan, Fuchs, and Jordan.  Such efforts spanning two centuries culminated in the famous Beukers-Heckman classification \cite{beukers1989monodromy}.\footnote{Their classification made key use of the classification of finite irreducible reflection groups by \cite{shephard1954finite}.
}
Using the Beukers-Heckman result, Franc and Mason \cite{franc2020classification} classified some three-dimensional RCFTs---namely those with exactly three modules (the Mathur-Mukhi-Sen classification) and which solve $d=3$ MDEs with irreducible monodromy. In this section, we give the full classification.

\subsubsection*{Hypergeometric equation}
Begin by considering the generic order-3 hypergeometric equation,
\bea
x \prod_{i=1}^3\left(x {d \over d x} +a_i\right) f = \prod_{i=1}^3\left(x {d \over d x} +b_{i+1}-1\right) f
\eea
for parameters $a_i,b_i \in \CC$.  Since this equation has three singular points, the monodromy group is generated by $M_0, M_1, M_\infty$, with corresponding eigenvalues
\bea
\label{eq:3dmoneigenvals}
&M_0:& \hspace{0.5 in} 1, \,\, e^{2 \pi i (1-b_1)}, \,\, e^{2 \pi i (1-b_2)}\no
\\
&M_1:& \hspace{0.5 in}1, \,\, 1, \,\, e^{2 \pi i (b_1+b_2-a_1-a_2-a_3)}\no
\\
&M_\infty:& \hspace{0.5 in}e^{2 \pi i a_1}, \,\,e^{2 \pi i a_2}, \,\,e^{2 \pi i a_3}
\eea
and the {monodromy group} is given by
\bea
\cM(a_i,b_i) = \langle M_0, M_1, M_\infty  \,| \, \,M_0 M_1 M_\infty =1 \rangle \subseteq \GL(3, \CC) ~,
\eea
c.f. (\ref{eq:mongroupdef}).

We will be interested in hypergeometric equations which admit a basis of algebraic solutions, since it is these which will give rise to integral characters. It is known that a differential equation admits such a basis if and only if the monodromy group is finite \cite{gray2008linear}, which occurs only for certain values of $(a_i, b_i)$. We are thus led to consider finite subgroups of $\GL(3,\CC)$.

The finite subgroups of $\GL(3,\CC)$ can be obtained by first identifying the subgroups of $\SL(3, \CC)$. These in turn are usefully organized in terms of \textit{transitivity} and \textit{primitivity}. Since these concepts will reappear throughout this section, we begin with a quick review of them.  Consider a vector space $V$ of dimension $d$ and a group $G \subset \GL(d, \CC)$. The group $G$ is called \textit{intransitive} if there is a direct sum decomposition $V = V_1 \oplus V_2 \oplus \dots \oplus V_n$ with ${\rm dim}\,V_j \geq 1$ and $n \geq 2$ such that $G$ leaves the spaces $V_j$ invariant. If no such decomposition exists, the group $G$ is called \textit{transitive}. A transitive group $G$ is called \textit{imprimitive} if there is a direct sum decomposition $V = V_1 \oplus V_2 \oplus \dots \oplus V_n$ with ${\rm dim}\,V_j \geq 1$ and $n \geq 2$ such that $G$ permutes the spaces $V_j$. If such a decomposition does not exist, $G$ is called primitive. These concepts can be illustrated schematically as follows,
 \bea
& \mathrm{Intransitive}:& \left(\begin{matrix} \times & 0 & 0 \\ 0 & \times & 0 \\ 0 & 0 & \times \end{matrix} \right), \left(\begin{matrix} \times & 0 & 0 \\ 0 & \times & \times \\ 0 & \times & \times \end{matrix} \right)
 \no\\
 & \mathrm{Imprimitive}:& \left(\begin{matrix} 0 & 0 & \times \\ \times & \times & 0 \\ \times & \times & 0 \end{matrix} \right), \left(\begin{matrix} 0 & \times & \times \\ 0 & \times & \times \\ \times & 0 & 0 \end{matrix} \right)\no
 \\
  & \mathrm{Primitive}:& \left(\begin{matrix} \times & \times & \times \\ \times & \times & \times \\ \times & \times & \times \end{matrix} \right)
 \eea
Intransitive subgroups are decomposable, and this means that the tentative three-character theory actually reduces to a product of one- and two-character theories. We thus neglect intransitive subgroups in what follows.

The transitive imprimitive and transitive primitive finite subgroups of $\SL(3,\CC)$ were described by Blichfeldt, Dickson, and Miller in \cite{miller1916theory}, and independently in the physics literature by Fairbairn, Fulton, and Klink \cite{Fairbairn:1964sga}. The original lists have some minor omissions, with the corrected lists given in \cite{yau1993gorenstein,Hanany:1998sd}. These results are reviewed in Appendix \ref{app:SL3subs}: to summarize, there are eight transitive primitive subgroups (the analogs of the three finite exceptional subgroups of $\SL(2,\CC)$) and two infinite series of transitive imprimitive subgroups (the analogs of the dihedral series of $\SL(2,\CC)$). We denote them as follows,
\bea
\mathrm{Transitive \,\, primitive}&:& \hspace{0.5 in} G_I^{(3)}, \dots, G_{VIII}^{(3)}
\no\\
\mathrm{Transitive \,\, imprimitive}&:&\hspace{0.5 in} G^{(3)}(a,b), \,\,G^{(3)}(a,b,a',b')
\eea
with all definitions given in Appendix \ref{app:SL3subs}. Notable cases include $G_{II}^{(3)}$, a group of order 216 known as the \textit{Hesse group}; $G_{IV}^{(3)}$, which is isomorphic to the icosahedral group $A_5$; $G_{VI}^{(3)}$, a group of order 168 known as the \textit{Klein group}; and $G_{VII}^{(3)}$, a group of order 1080 known as the \textit{Valentiner group}.

Each of the above finite subgroups gives a subgroup in $\GL(3, \CC)$ upon appropriate extension. Fortunately, we will not have to concern ourselves with the analysis of these extensions. Instead, we may make use of the work of Beukers and Heckman \cite{beukers1989monodromy}. In the current context, the results of Beukers and Heckman states that if the monodromy group $\cM(a_i, b_i)$ of a hypergeometric equation is primitive, then either $\cM(a_i, b_i)$, $\cM\left(a_i+ {1\over 3}, b_i+ {1\over 3}\right)$, or $\cM\left(a_i+ {2\over 3}, b_i+ {2\over 3}\right)$ is a \textit{complex reflection group}.\footnote{Technically the statement of \cite{beukers1989monodromy} is that if the \textit{reflection subgroup} $\cM_r(a_i, b_i)$ of $\cM(a_i, b_i)$ is primitive, then $\cM(a_i+\delta, b_i+\delta)$ is isomorphic to  $\cM_r(a_i, b_i)$ for some $\delta \in \mathbb{Q}$. In our case we consider only $\delta \in \left\{0, {1\over3},{2 \over 3} \right\}$ since otherwise (\ref{eq:3dl0fixed}) will be violated. This technicality will not be relevant for our purposes} Hence for the primitive cases it suffices to classify three-dimensional complex reflection groups. In fact, the set of such groups is known and is quite limited, as shown in Table \ref{tab:comprefgroups}.  This makes the classification program for primitives much simpler than may have otherwise been expected. For imprimitive subgroups, we will simply work projectively.

\begin{table}[!tp]
\begin{center}
\begin{tabular}{c|c|c|c}
& group & order & center
\no\\\hline
$W(H_3)$ & $\ZZ_2 \,\Bowtie\, G_{IV}^{(3)}$ & 120 & 2
\no\\
$W(J_3(4))$ & $\ZZ_2 \,\Bowtie\ G_{VI}^{(3)}$ & 336 & 2
\no\\
$W(L_3)$ & $G_I^{(3)}$& 648 & 3
\no\\
$W(M_3)$ &$ \ZZ_2 \,\Bowtie\ G_I^{(3)}$ & 1296 & 6
\no\\
$W(J_3(5))$ & $\ZZ_2 \,\Bowtie\ G_{VIII}^{(3)}$ & 2160 & 6
\end{tabular}
\end{center}
\caption{Three-dimensional complex reflection groups}
\label{tab:comprefgroups}
\end{table}

Before applying these results to degree-3 MDEs, let us briefly summarize the contents of Table \ref{tab:comprefgroups}. First, we see that the product of the icosahedral group $G_{IV}^{(3)}$ with a group of order 2 gives a complex reflection group of order 120. Likewise, the product of the Klein group $G_{VI}^{(3)}$ with a group of order 2 gives a complex reflection group of order 336.  The triple cover of the Hesse group $G_{II}^{(3)}$, namely $G_{I}^{(3)}$, is a complex reflection group of order 648. The product of this group with a group of order two gives rise to another reflection group of order 1296. Finally, the product of the Valentiner group $G_{VIII}^{(3)}$ with a group of order 2 gives a complex reflection group of order 2160.

\subsubsection*{Third-order MDE}
We now apply the results of the previous subsection to the study of three-character RCFTs. We begin by making the change of variables $x =1728^{-1} j(\tau)$. A tedious but conceptually straightforward computation gives
\bea
&\hspace{-0.2cm}\!\!\left[ D^{(3)} \! +\! \left(\!\m_1 {E_4^2 \over E_6} \!+\! \m_2 {E_6 \over E_4} \right)\!D^{(2)}
\!\!+\! \left(\half \m_1 {E_4^4 \over E_6^2} \!+\! \m_3 {E_6^2 \over E_4^2 }\!+\! \g_1 E_4 \right)\!D^{(1)} \!\!+\! \m_4 {E_6^3 \over E_4^3} \!+\! \g_2 E_6\right] \!\chi = 0~,
\eea
where the parameters are identified as follows,
\bea
\m_1 &=& \sigma_1(b) - \sigma_1(a) - {3\over 2}~,
\no\\
\m_2 &=&2- \sigma_1(b)~,
\no\\
\m_3 &=& {1 \over 9}\left(19- 15 \sigma_1(b) + 9 \sigma_2(b) \right)~,
\no\\
\m_4 &=& - \sigma_3(b-1)~,
\no\\
\g_1 &=& {1 \over 6}\left(2 \sigma_1(a) - 6 \sigma_2(a) -5 \sigma_1(b) + 6 \sigma_2(b) +4 \right)~,
\no\\
\g_2 &=&\sigma_3(b-1) - \sigma_3(a)~.
\eea
Here we denote by $\sigma_p$ the degree-$p$ symmetric function in three variables, e.g. $\sigma_1(a) = a_1 + a_2 + a_3$.
One set of solutions is given by
\bea
\label{eq:3dMDEsols}
\chi_1(\tau) &=& j(\tau)^{-a_1}\, {}_3 F_2\left(d_{a_1,b_1},d_{a_1,b_2}, d_{a_1,b_3}\,;\, d_{a_1,a_2}, d_{a_1,a_3}\,;\, 1728 j(\tau)^{-1} \right) ~,
\no\\
\chi_2(\tau) &=&j(\tau)^{-a_2} \,{}_3 F_2\left(d_{a_2,b_1},d_{a_2,b_2}, d_{a_2,b_3}\,;\, d_{a_2,a_1}, d_{a_2,a_3}\,;\, 1728 j(\tau)^{-1} \right) ~,
\no\\
\chi_3(\tau) &=&j(\tau)^{-a_3} \,{}_3 F_2\left(d_{a_3,b_1},d_{a_3,b_2}, d_{a_3,b_3}\,;\,  d_{a_3,a_1}, d_{a_3,a_2}\,;\, 1728 j(\tau)^{-1} \right) ~,
\eea
where $d_{x,y} := 1+x-y$.
In the $\ell=0$ case, which will be the situation of most interest to us here, we require that $\m_1 = \m_2 = \m_3 = 0$, which in turn requires
\bea
\label{eq:3dl0fixed}
\left\{b_1,b_2, b_3\right\} = \left\{1,{1\over 3}, {2\over 3}\right\}\,\,{\rm mod}\,\,1~,\hspace{0.5 in} a_1 + a_2 + a_3 = \half~.
\eea
By (\ref{eq:3dl0fixed}) and (\ref{eq:3dmoneigenvals}), we require that for $\ell=0$ we have
\bea
\label{eq:3dSSTdettrac}
\mathrm{det}\,S = -1~, \hspace{0.5 in} \mathrm{Tr}\,S= 1~, \hspace{0.5 in} \mathrm{det}\,ST = 1~, \hspace{0.5 in} \mathrm{Tr}\,ST= 0~.
\eea
These facts, together with the facts that $S$ must be an element of order 2 and $ST$ and element of order 3, will be enough to fix all allowed combinations of $c, h_1,$ and $h_2$.

\subsubsection*{Classification}
\paragraph{Primitive Monodromy}

\begin{table}[!tp]
\begin{center}
\begin{tabular}{c|cccccccccc}
&$1a$& $2a$& $2b$ & $2c$& $3a$& $5a$& $5b$ & $6a$& $10a$& $10b$
\no\\\hline
$\chiG_1$ & 1& 1& 1& 1& 1& 1& 1& 1& 1& 1
\no\\
$\chiG_2$ & 1& $-1$& 1&$ -1$& 1& 1& 1& $-1$& $-1$& $-1$
\no\\
$\chiG_3$ & 3&$ -3$&$ -1$& 1& 0& $ \phi_+$&$ \phi_-$& 0& $- \phi_+$& $- \phi_-$
\no\\
$\chiG_4$ & 3& 3& $-1$& $-1$& 0& $ \phi_-$& $ \phi_+$& 0&$  \phi_-$& $ \phi_+$
\no\\
$\chiG_5$ & 3& 3& $-1$& $-1$& 0& $ \phi_+$& $ \phi_-$& 0& $ \phi_+$&$  \phi_-$
\no\\
$\chiG_6$& 3& $-3$& $-1$& 1& 0& $ \phi_-$& $ \phi_+$& 0& $- \phi_-$& $- \phi_+$
\no\\
$\chiG_7$ & 4& 4& 0& 0& 1& $-1$& $-1$& 1& $-1$& $-1$
\no\\
$\chiG_8$ & 4& $-4$& 0& 0& 1& $-1$& $-1$& $-1$& 1& 1
\no\\
$\chiG_9$ & 5& 5& 1& 1& $-1$& 0& 0&$ -1$& 0& 0
\no\\
$\chiG_{10}$ & 5& $-5$& 1& $-1$& $-1$& 0& 0& 1& 0& 0
\end{tabular}
\end{center}
\caption{Character table of $W(H_3)$. We define $ \phi_\pm = {1\over 2}(1\pm \sqrt{5})$.}
\label{tab:WH3chartab}
\end{table}
\noindent
Consider the complex reflection group $W(H_3)$, which has character table shown in Table \ref{tab:WH3chartab}. We see that there are four three-dimensional representations of this group, with corresponding group characters $\chiG_3,\dots, \chiG_6$. Only the two faithful representations $\chiG_3$ and $\chiG_6$ have an order $2$ conjugacy class with trace 1 which can play the role of $S$. In both cases $S$ must be located in the class $2c$ and $ST$ in the class $3a$. This means that $T$ must be in one of the classes appearing in the product (without multiplicities),\footnote{Note that these fusion rules can be read off entirely from the character table, using an analog of the usual Verlinde formula. In the current case, we need (a subset of) the Verlinde formula for $\mathrm{Rep}(Z[G])$, the representation category of the Drinfeld center of the fusion category defined by $G$,
\bea
N_{ab}^c = {|G|\over |C_G(a)| |C_G(b)| }\sum_{i \in \mathrm{irreps}}{\chiG_i(a) \chiG_i(b) \overline{\chiG_i}(c)\over \chiG_i(1)}~,
\eea
where $|G|$ is the order of the group and $|C_G(a)|$ is the number of elements in the conjugacy class $a$.
This gives the multiplicity in the fusion channel $a \otimes b \rightarrow c$.}
\bea
2c \otimes 3 a = 2c\oplus 6a\oplus 10a \oplus 10b~.
\eea
Then using $\mathrm{det}(T)=\mathrm{det}(S)\mathrm{det}(ST) = -1$ and the values of $\mathrm{Tr}\,T$ given in the character table, we are able to conclude that the eigenvalues of $T$ are of the form $e^{2 \pi i a_i}$ where the exponents $\left\{a_1, a_2, a_3 \right\}$ take the following values,
\bea
&6a:&  \hspace{0.5in} \left\{{1\over 6}, {1\over 2}, {5 \over 6} \right\}\,\,\,\mathrm{mod}\,\,1~,
\no\\
&10a,b:& \hspace{0.5in} \left\{{1\over 10}, {1\over 2}, {9 \over 10} \right\},  \,\, \left\{{3\over 10}, {1\over 2}, {7 \over 10} \right\}\,\,\,\mathrm{mod}\,\,1~.
\eea
c.f. Table \ref{tab:3dl3}. This then gives us all possible values of $(c,h_1,h_2)$. For example considering the class $\left\{{1\over 10}, {1\over 2}, {9 \over 10} \right\}$, we have the following possibilities: for $m,n \in {1\over 3}\ZZ$,
\bea
&\mathrm{I}.& \hspace{0.2 in}-{c \over 24 } = {1\over 10} + m , \hspace{0.5 in} h_1 -{c \over 24 } = {1\over 2}+n , \hspace{0.57 in} h_2 -{c \over 24 } = {9\over 10}-m-n-1
\no\\
&\mathrm{II}.&\hspace{0.2 in}-{c \over 24 } = {1\over 2} + m , \hspace{0.58 in} h_1 -{c \over 24 } = {1\over 10}+n , \hspace{0.5 in} h_2 -{c \over 24 } = {9\over 10}-m-n-1
\no\\
&\mathrm{III}.&\hspace{0.2 in}-{c \over 24 } = {9\over 10} + m , \hspace{0.5 in} h_1 -{c \over 24 } = {1\over 10}+n , \hspace{0.5 in} h_2 -{c \over 24 } = {1\over 2}-m-n-1
\no
\eea
Note that we are allowing for shifts in $\ZZ/3$ since by Beukers-Heckman, any one of $\cM(a_i, b_i)$, $\cM\left(a_i+ {1\over 3}, b_i+ {1\over 3}\right)$, or $\cM\left(a_i+ {2\over 3}, b_i+ {2\over 3}\right)$ can be identified with $W(H_3)$.

  \begin{table}[!tp]
\begin{minipage}[t]{.33\linewidth}
\centering
\begin{tabular}[t]{c|c|c}
\multicolumn{3}{c}{$\,\,W(H_3)$}
\\
 $c$ & $h_i$&$n_J$
\\\hline
 $4$ & ${2 \over5}$ , $3 \over 5$&24
 \\
  ${12}$ & ${3 \over 5}$ , $7 \over 5$&222
 \\
 ${20}$ & ${7 \over 5}$ , $8 \over 5$& 120
\end{tabular}\end{minipage}%
 \begin{minipage}[t]{.33\linewidth}
\centering
\begin{tabular}[t]{c|c|c}
\multicolumn{3}{c}{$\,\,W(H_3)$}
\\
 $c$ & $h_i$ &$n_J$
\\\hline
 ${4 \over 5}$ & ${1 \over 5}$ , $2 \over 5$ & 2
 \\
 ${12 \over 5}$ & ${1 \over 5}$ , $3 \over 5$  &3
\\
  ${28 \over 5}$ & ${2 \over 5}$ , $4 \over 5$&28
  \\
 ${36 \over 5}$ & ${3 \over5}$ , $4 \over 5$& 144
\\
  ${44 \over 5}$ & ${2 \over 5}$ , $6 \over 5$&220
  \\
  ${52 \over 5}$ & ${3 \over 5}$ , $6 \over 5$&104
 \\
  ${68 \over 5}$ & ${4 \over 5}$ , $7 \over 5$& 136
  \\
  ${76 \over 5}$ & ${3 \over 5}$ , $9 \over 5$&437
\\
 ${76 \over 5}$ & ${4 \over 5}$ , $8 \over 5$&380
\\
${84 \over 5}$ & ${6 \over 5}$ , $7 \over 5$&336
 \\
  ${92 \over 5}$ & ${6 \over 5}$ , $8 \over 5$& 92
\\
 ${108 \over 5}$ & ${4 \over 5}$ , $12 \over 5$& 1404
\\
 ${108 \over 5}$ & ${7 \over 5}$ , $9 \over 5$ & 27
\\
 ${116 \over 5}$ & ${8 \over 5}$ , $9 \over 5$ &58
\\
 ${164 \over 5}$ & ${11\over 5}$ , $12 \over 5$&0
\end{tabular}
  \end{minipage}
   \begin{minipage}[t]{.33\linewidth}
\centering
\begin{tabular}[t]{c|c|c}
\multicolumn{3}{c}{$\,\,W(J_3(4))$}
\\
 $c$ & $h_i$& $n_J$
\\\hline
 ${4 \over 7}$ & ${1 \over 7}$ , $3 \over 7$ & 1
\\
 ${12 \over 7}$ & ${2 \over 7}$ , $3 \over 7$& 6
\\
  ${44 \over 7}$ & ${4 \over 7}$ , $5 \over 7$& 88
\\
  ${52 \over 7}$ & ${4 \over 7}$ , $6 \over 7$& 156
\\
  ${60 \over 7}$ & ${3 \over 7}$ , $8 \over 7$& 210
\\
 ${68 \over 7}$ & ${3 \over 7}$ , $9 \over 7$& 221
\\
  ${100 \over 7}$ & ${4 \over 7}$ , $12 \over 7$& 380
\\
 ${100 \over 7}$ & ${5 \over 7}$ , $11 \over 7$& 325
\\
 ${108 \over 7}$ & ${6 \over 7}$ , $11 \over 7$& 378
\\
 ${108 \over 7}$ & ${4 \over 7}$ , $13 \over 7$&456
\\
 ${116 \over 7}$ & ${8 \over 7}$ , $10 \over 7$&348
\\
 ${124 \over 7}$ & ${9 \over 7}$ , $10 \over 7$& 248
\\
 ${156 \over 7}$ & ${11 \over 7}$ , $12 \over 7$& 78
\\
 ${156 \over 7}$ & ${5 \over 7}$ , $18 \over 7$ &1248
\\
 ${164 \over 7}$ & ${11 \over 7}$ , $13 \over 7$ & 41
\\
 ${236 \over 7}$ & ${16 \over 7}$ , $17 \over 7$ & 0
\end{tabular}
  \end{minipage}
   \caption{Possible three-character $\ell=0$ theories with primitive monodromy group. Compare with Table \ref{tab:3dl0tabimprim1}.}
  \label{tab:3dl0tab}
\end{table}

 Now what remains is to check which such choices of $(c,h_1,h_2)$ give rise to characters with a vacuum and completely positive Fourier coefficients. As usual we simply turn to an explicit computerized search, which has already been carried out above. The results as shown in the first two columns of Table \ref{tab:3dl0tab}. This completes the classification of theories with monodromy descending from $W(H_3)$. The steps outlined above may be repeated for the remaining primitive monodromy groups, with the results given in Table \ref{tab:3dl0tab}. We see that besides $W(H_3)$, only $W(J_3(4))$ gives rise to physically sensible characters.

 Comparing to Table \ref{tab:3dl0tabimprim1}, we see that we have reproduced all cases involving denominators $5$ and $7$. This allows us to not only check this portion of our results, but also to identify the precise monodromy groups acting on these characters. The remaining entries of Table \ref{tab:3dl0tabimprim1} must have imprimitive monodromy group.

\paragraph{Imprimitive Monodromy} We now move on to a classification of theories with imprimitive monodromy. This portion of the classification has already been studied in \cite{mathur1989group}, so we will be brief. Recall that in $\SL(3,\CC)$, there were two infinite families of imprimitive finite groups $G^{(3)}(a,b)$ and $G^{(3)}(a,b,a',b')$, which are explicitly
\bea
G^{(3)}(a,b) &:=& \left\langle A, B \right\rangle~,\hspace{0.75 in} A:=\left(\begin{matrix} a & 0 & 0 \\ 0 & b & 0 \\ 0 & 0 & {1 \over a b}\end{matrix} \right),\, B:= \left(\begin{matrix} 0 & 1 & 0 \\ 0 &0 & 1 \\ 1 &0 & 0  \end{matrix} \right)~,\no
\\
G^{(3)}(a,b,a',b')&:=& \left\langle A, B,C \right\rangle~,\hspace{0.5 in} C:=\left(\begin{matrix} 0 & a'& 0 \\ b' & 0 & 0 \\ 0 & 0 & -{1 \over a' b'}\end{matrix} \right)
\eea
By the analysis of \cite{mathur1989group}, it can be shown that the allowed set of $(c,h_1, h_2)$ in theories with these monodromies are
\bea
G^{(3)}(a,b):& \hspace{0.2 in}&(c, h_1, h_2) = \left(4 + 8 (m+n), \, {1\over 3} + m, \, {2\over 3}+ n\right)~,\hspace{0.5 in} m,n\in \ZZ~,
\no\\
G^{(3)}(a,b,a',b'):& \hspace{0.2 in}&(c,h_1,h_2 ) = \left( 2 h_1 + k, \, h_1, \, h_1 + {2 k + 1 \over 2} \right)~, \hspace{0.5 in } h_1 \in \mathbb{Q}, k \in \ZZ~,\no
\eea
where $m,n$ appearing here are related to $a,b$ in some appropriate way (the details of which are unimportant for us), and likewise for $k$ and $a,b,a',b'$.
One can now simply scan over all values to identify which cases have positive characters with a vacuum. Alternatively, one could identify the legitimate cases by simply comparing to Table \ref{tab:3dl0tabimprim1}. We find that all cases with a 3 denominator in Table \ref{tab:3dl0tabimprim1} have monodromy group of the type $G^{(3)}(a,b)$, while the remaining cases, including the infinite family of $\rm{Spin}(n)_1$ theories which are not included in Table \ref{tab:3dl0tabimprim1}, all have monodromy group of the type $G^{(3)}(a,b,a',b')$.

\subsection{Four characters}

\label{Sec:d=4}
\subsubsection*{Fourth-order MDE}
We now provide a similar analysis for the four-character case. The main qualitative difference between the three- and four-character cases is the fact that the relevant MDE in the latter case cannot be expressed in terms of a hypergeometric equation. Indeed, the degree-four monic MDE is given by
\ie
\left[ D^{(4)} + \m_1 E_4 D^{(2)} + \m_2 E_6 D^{(1)} + \m_3 E_4^2 \right] \chi = 0
\fe
and upon making the usual change of variables $x = 1728^{-1}j(\tau)$ we obtain the following differential equation,
\ie
&\left[ \left( x \frac{d}{dx} \right)^4 - \frac{x+2}{1-x} \left( x \frac{d}{dx} \right)^3 + \frac{(36 \m_1+11) x^2-4 (9 \m_1+7) x+44}{36 (x-1)^2} \left( x \frac{d}{dx} \right)^2\right.
\\
& \hspace{.6in} \left.- \frac{x^2 (-6 \m_1+36 \m_2-1)-4 x (3 \m_1+9 \m_2+1)+8}{36 (x-1)^2} x \frac{d}{dx} + \frac{\m_3 x^2}{(x-1)^2} \,\right]f(x) = 0
\fe
which is not of hypergeometric type. Nevertheless, this differential equation again has three singular points at $x =  0, \, 1, \,\infty$, which are the images of $\tau = \omega,\, i ,\, i \infty$. As such, the monodromy group is again of the form (\ref{eq:mongroupdef}).

In the three-character case, the monodromy group was \textit{a priori} a generic finite subgroup of $\GL(3,\CC)$. However, thanks to the results of Beukers-Heckman we were able to reduce the analysis to three-dimensional complex reflection groups, of which there were few. We might again hope to leverage these results to avoid a classification of generic finite subgroups of $\GL(4,\CC)$, but unfortunately the results of Beukers and Heckman are only applicable to the case of hypergeometric equations, and thus there is no such reduction to complex reflection groups in the current case.

It would thus seem that in order to understand all possible monodromy groups of the monic degree-4 MDE, we would have to consider all finite subgroups of $\GL(4,\CC)$. The situation is slightly better than that though, and in fact we will only have to consider finite subgroups of $\SL(4,\CC)$. To see why, begin by denoting the exponents at $x=0,1,$ and $\infty$ by $\{b_1, \dots, b_4\}, \{c_1, \dots, c_4\},$ and $\{a_1,\dots, a_4\}$.\footnote{By this we mean that around $x=\infty$ the four solutions admit a series expansion of the form $f_i(x) \sim x^{-a_i} + \dots $, and likewise for the other points.} Then by solving the indicial equation at $x = 0, \, 1$, we find solutions
\ie
\label{d=4Betas}
\left\{ b_1, b_2, b_3, b_4 \right\} = \left\{ 0, \, \frac13, \, \frac23, \, 1 \right\} \mod 1 \, ,
\fe
and
\ie
\label{d=4Gammas}
\left\{ c_1, c_2, c_3, c_4 \right\} = \left\{ 0, \, \frac12, \, 1, \, \frac32 \right\} \mod 1 \, ,
\fe
insensitive to the coefficients $\m_1, \m_2, \m_3$.  Therefore, the monodromy matrices $S$ and $ST$ satisfy
\ie
\label{eq:4dSSTdettrac}
\Tr S = 0 \, , \hspace{0.5 in} \det S = 1 \, , \hspace{0.5 in} \Tr ST = 1 \, , \hspace{0.5 in} \det ST = 1 \, .
\fe
These are the degree-four analogs of (\ref{eq:3dSSTdettrac}). Crucially however, in the current case we see that both $S$ and $ST$---and hence every matrix in the group that they generate---is of determinant one. For this reason the monodromy group is a subgroup of $\SL(4,\CC)$.

The classification of finite subgroups of $\SL(4, \CC)$ was initiated by Blichfeldt \cite{blichfeldt1917finite} and later completed by Hanany and He \cite{Hanany:1999sp}.  For completeness, we summarize the results of \cite{Hanany:1999sp} in Appendix \ref{app:SL4subs}. As in the three-character case, the classification may be split into  transitive primitive and transitive imprimitive cases.
There are now thirty transitive primitive subgroups (the analogs of the three exceptional subgroups of $\SL(2,\CC)$) and five infinite series of transitive imprimitive subgroups (the analogs of the dihedral series of $\SL(2,\CC)$). We denote them as follows,
\bea
\label{eq:SL4Csublist}
\mathrm{Transitive \,\, primitive}&:& \hspace{0.5 in} G_I^{(4)},\, \dots,\, G_{XXX}^{(4)}
\no\\
\mathrm{Transitive \,\, imprimitive}&:&\hspace{0.5 in} G_1^{(4)}(n),\, \dots,\, G_4^{(4)}(n),\, G^{(4)}(n;a,b,c,d,e,f)
\eea
with all definitions given in Appendix \ref{app:SL4subs}.

\subsubsection*{Classification}

Having introduced the relevant candidate monodromy groups, we may now apply the techniques developed in the three-character case.
As usual we divide the discussion into primitive and imprimitive monodromy.

\paragraph{Primitive monodromy}  We begin with the monodromy group $G_I^{(4)}$, which is isomorphic to a double cover of $\SL(2, 5)$.  The character table of this group is given in Table \ref{Tab:CharacterTable}. We see that the monodromy $S$ matrix must belong to class $2b$, and $ST$ must belong to class $3a$.  Together they should generate one of the two faithful four-dimensional irreducible representations.  The possible conjugacy classes for $T$ are constrained by the class fusion (ignoring multiplicity),
\ie
2b \otimes 3a = 12a \oplus 20a \oplus 20b \oplus 20c \oplus 12b \oplus 20d \oplus 2b \, .
\fe
According to the character table, the order and trace of $T$, from which the possible sets of exponents can be deduced, fall into the following cases:
\begin{itemize}
\item $T^{12} = 1 \, , ~ \Tr T = \pm i$~:~
\ie
\left\{a_1, \dots, a_4 \right\} = \pm \left\{ 0, \frac{1}{12}, \frac{5}{12}, \frac{1}{2} \right\} \, , ~ \pm \left\{ \frac14, \frac14, \frac{7}{12}, \frac{11}{2} \right\} \mod 1 \, .
\fe
\item $T^{20} = 1 \, , ~ \Tr T = \pm i$~:~
\ie
\left\{a_1, \dots, a_4 \right\} = \pm \left\{\frac{3}{20},\frac{7}{20},\frac{11}{20},\frac{19}{20}\right\} \mod 1 \, .
\fe
\item $T^2 = 1 \, , ~ \Tr T = 0$~:~
\ie
\left\{a_1, \dots, a_4 \right\} = \left\{0, 0, \frac12, \frac12\right\} \mod 1 \, .
\fe
\end{itemize}
Repeating the above analysis for all primitive subgroups of $\SL(4, \CC)$, we obtain the full set of possible sets of exponents mod 1,.
As usual, one may then check which of these cases gives rise to a unique vacuum and positive Fourier coefficients. Repeating this exercise for all primitive finite subgroups reproduces the first four columns of Table \ref{Tab:Unitary}. The final column must then correspond to cases with imprimitive monodromy.

\begin{table}[htp!]
\centering
\scalebox{1}{
$
\begin{array}{c|cccccccccccccccccc}
 & {1a} & {6a} & {5a} & {5b} & {10a} & {10b} &{3a} & \text{2a} & {4a} & {4b} & {12a} & {20a} & {20b} & 20 & {12b} & {20d} & 4 & {2b} \\\hline
 \chiG_1 & 1 & 1 & 1 & 1 & 1 & 1 & 1 & 1 & 1 & 1 & 1 & 1 & 1 & 1 & 1 & 1 & 1 & 1 \\
 \chiG_2 & 1 & 1 & 1 & 1 & 1 & 1 & 1 & 1 & 1 & -1 & -1 & -1 & -1 & -1 & -1 & -1 & -1 & -1 \\
 \chiG_3 & 2 & 1 & -\phi_+ & -\phi_- & \phi_+ & \phi_- & -1 & -2 & 0 & -2 i & i & -i \phi_+ & -i \phi_- & i \phi_- & -i & i \phi_+ & 2 i & 0 \\
 \chiG_4 & 2 & 1 & -\phi_- & -\phi_+ & \phi_- & \phi_+ & -1 & -2 & 0 & -2 i & i & -i \phi_- & -i \phi_+ & i \phi_+ & -i & i \phi_- & 2 i & 0 \\
 \chiG_5 & 2 & 1 & -\phi_+ & -\phi_- & \phi_+ & \phi_- & -1 & -2 & 0 & 2 i & -i & i \phi_+ & i \phi_- & -i \phi_- & i & -i \phi_+ & -2 i & 0 \\
 \chiG_6 & 2 & 1 & -\phi_- & -\phi_+ & \phi_- & \phi_+ & -1 & -2 & 0 & 2 i & -i & i \phi_- & i \phi_+ & -i \phi_+ & i & -i \phi_- & -2 i & \
0 \\
 \chiG_7 & 3 & 0 & \phi_- & \phi_+ & \phi_- & \phi_+ & 0 & 3 & -1 & 3 & 0 & \phi_- & \phi_+ & \phi_+ & 0 & \phi_- & 3 & -1 \\
 \chiG_8 & 3 & 0 & \phi_+ & \phi_- & \phi_+ & \phi_- & 0 & 3 & -1 & 3 & 0 & \phi_+ & \phi_- & \phi_- & 0 & \phi_+ & 3 & -1 \\
 \chiG_9 & 3 & 0 & \phi_- & \phi_+ & \phi_- & \phi_+ & 0 & 3 & -1 & -3 & 0 & -\phi_- & -\phi_+ & -\phi_+ & 0 & -\phi_- & -3 & 1 \\
 \chiG_{10} & 3 & 0 & \phi_+ & \phi_- & \phi_+ & \phi_- & 0 & 3 & -1 & -3 & 0 & -\phi_+ & -\phi_- & -\phi_- & 0 & -\phi_+ & -3 & 1 \\
 \chiG_{11} & 4 & 1 & -1 & -1 & -1 & -1 & 1 & 4 & 0 & 4 & 1 & -1 & -1 & -1 & 1 & -1 & 4 & 0 \\
 \chiG_{12} & 4 & 1 & -1 & -1 & -1 & -1 & 1 & 4 & 0 & -4 & -1 & 1 & 1 & 1 & -1 & 1 & -4 & 0 \\
 \chiG_{13} & 4 & -1 & -1 & -1 & 1 & 1 & 1 & -4 & 0 & -4 i & -i & -i & -i & i & i & i & 4 i & 0 \\
 \chiG_{14} & 4 & -1 & -1 & -1 & 1 & 1 & 1 & -4 & 0 & 4 i & i & i & i & -i & -i & -i & -4 i & 0 \\
 \chiG_{15} & 5 & -1 & 0 & 0 & 0 & 0 & -1 & 5 & 1 & 5 & -1 & 0 & 0 & 0 & -1 & 0 & 5 & 1 \\
 \chiG_{16} & 5 & -1 & 0 & 0 & 0 & 0 & -1 & 5 & 1 & -5 & 1 & 0 & 0 & 0 & 1 & 0 & -5 & -1 \\
 \chiG_{17} & 6 & 0 & 1 & 1 & -1 & -1 & 0 & -6 & 0 & -6 i & 0 & i & i & -i & 0 & -i & 6 i & 0 \\
 \chiG_{18} & 6 & 0 & 1 & 1 & -1 & -1 & 0 & -6 & 0 & 6 i & 0 & -i & -i & i & 0 & i & -6 i & 0 \\
\end{array}
$
}
\caption{The character table of a primitive finite subgroup $G_I^{(4)}$ of $\SL(4, \CC)$. We define $ \phi_\pm = {1\over 2}(1\pm \sqrt{5})$.}
\label{Tab:CharacterTable}
\end{table}

\paragraph{Imprimitive monodromy} We begin the classification of theories with imprimitive monodromy by starting with $G^{(4)}(n;a,b,c,d,e,f)$. For convenience we recall the definition of this group here,
\bea
G^{(4)}(n;a,b,c,d,e,f) = \left \langle A, B_1, \dots, B_n \right\rangle~,
\eea
where we have
\bea
A:=  \begin{pmatrix} a & b & 0 & 0 \\ c & {1+b c \over a}& 0 & 0 \\ 0 & 0 & d & e \\ 0 & 0 & f &  {1+ef\over d} \end{pmatrix} , \hspace{0.2 in}B_i:= \begin{pmatrix}0 & 0 & 1 & 0 \\ 0 & 0 & 0 & 1 \\ \xi_n^i & 0 & 0 & 0 \\ 0 & \xi_n^{-i} & 0 & 0  \end{pmatrix} ~.
\eea
Each element of $G^{(4)}(n;a,b,c,d,e,f)$ is in one of the following classes: $[A]$ (block diagonal matrices with $\SL(2,\CC)$ blocks), $[B_i]$, or $[AB_i]$ (off-diagonal matrices $\begin{pmatrix} 0 &  M \\ N & 0\end{pmatrix}$, again with $\SL(2,\CC)$ blocks). The second case is just a special case of the third.

We begin by noting that $[B_i]$ and $[AB_i]$ have even order, so $ST$ must be an element of $[A]$. Demanding that $(ST)^3=1$ and $\mathrm{Tr}\,ST =1$ then gives constraints on the particular matrix. If $S$ is also in $[A]$ then both of the generators are block diagonal, and we get an intransitive group. Hence we instead consider $S$ in $[B_i]$ or $[AB_i]$. Demanding that $S^2 =1$ and $\mathrm{Tr}\,S=0$ again gives constraints on the particular matrix. Finally, evaluating $T=S(ST)$ subject to the above constraints then allows us to read off the eigenvalues of $T$. We find a unique possibility (up to integer shifts): $\{{1\over 6}, {5\over 6}, {1\over 3}, {2\over 3}\}$.

A similar analysis can be carried out for the imprimitive subgroups of type $G_i^{(4)}(n)$ for $i=1, \dots, 4$, though it is significantly more tedious and hence we relegate the details to Appendix \ref{app:4dimprim}. The full set of possible exponents for these cases are given by
\bea
\label{eq:4dimprimres}
\left\{ {p \over n},\,\, -{p\over3n}, \,\,{2\over 3} - {p \over 3n},\,\, {1\over 3} - {p\over 3n} \right\}\hspace{0.5 in} p \in [-2n , 2n]
\eea
Searching for positive theories with a vacuum reproduces the right column of Table \ref{Tab:Unitary}.

\section*{Acknowledgements}

We are grateful to Eric Perlmutter for collaboration until the late stages of this project.  We thank Nathan Benjamin and Eric Perlmutter for comments on the first draft, and Yang-Hui He for sharing data on the character tables of finite subgroups of $\SL(3,\CC)$ and $\SL(4,\CC)$.  We also thank Ana Martínez Pastor for various group-theoretic consultations. J.P.-M. also thanks her for being his mother. This material is based upon work supported by the U.S. Department of Energy, Office of Science, Office of High Energy Physics, under Award Number DE-SC0011632.  YL is supported by the Simons Bootstrap Collaboration.

\appendix

\section{Lists of allowed exponents }
\label{app:extraexps}
In this appendix we catalogue all allowed exponents for (quasi)characters with $d\leq 5$ and any allowed $\ell$ (besides the ones which were already introduced in the main text).

\begin{table}[H]
\begin{center}
\scalebox{.9}{
\makebox[\textwidth]{
\begin{tabular}{c|c}
$n$ & 3d $\ell=0$ exponents
\\ \hline
4 &$ \left\{\frac{1}{4},\frac{1}{2},\frac{3}{4}\right\}$
\\
6&$\left\{\frac{1}{6},\frac{1}{2},\frac{5}{6}\right\}$
\\
8&$ \left\{\frac{1}{8},\frac{5}{8},\frac{3}{4}\right\},\left\{\frac{1}{4},\frac{3}{8},\frac{7}{8}\right\}$
\\
10&$\left\{\frac{1}{10},\frac{1}{2},\frac{9}{10}\right\}, \left\{\frac{3}{10},\frac{1}{2},\frac{7}{10}\right\}$
\\
12&$\left\{\frac{1}{12},\frac{7}{12},\frac{5}{6}\right\}, \left\{\frac{1}{6},\frac{5}{12},\frac{11}{12}\right\}$
 \\
 14&$\left\{\frac{1}{14},\frac{9}{14},\frac{11}{14}\right\},\left\{\frac{3}{14},\frac{5}{14},\frac{13}{14}\right\}$
 \\
16 & $\left\{\frac{1}{16},\frac{9}{16},\frac{7}{8}\right\},\left\{\frac{5}{16},\frac{3}{8},\frac{13}{16}\right\},\left\{\frac{3}{16},\frac{5}{8},\frac{11}{16}\right\},\left\{\frac{1}{8},\frac{7}{16},\frac{15}{16}\right\}$
\\
24&$\left\{\frac{1}{24},\frac{13}{24},\frac{11}{12}\right\}, \left\{\frac{1}{12},\frac{11}{24},\frac{23}{24}\right\},\left\{\frac{5}{24},\frac{7}{12},\frac{17}{24}\right\}, \left\{\frac{7}{24},\frac{5}{12},\frac{19}{24}\right\}$
\\
30 & $\left\{\frac{1}{30},\frac{19}{30},\frac{5}{6}\right\},\left\{\frac{7}{30},\frac{13}{30},\frac{5}{6}\right\},\left\{\frac{1}{6},\frac{11}{30},\frac{29}{30}\right\},\left\{\frac{1}{6},\frac{17}{30},\frac{23}{30}\right\}$
\\
42 & $\left\{\frac{1}{42},\frac{25}{42},\frac{37}{42}\right\},\left\{\frac{5}{42},\frac{17}{42},\frac{41}{42}\right\},\left\{\frac{11}{42},\frac{23}{42},\frac{29}{42}\right\},\left\{\frac{13}{42},\frac{19}{42},\frac{31}{42}\right\}$
\\
48 &$\left\{\frac{1}{48},\frac{25}{48},\frac{23}{24}\right\}, \left\{\frac{5}{48},\frac{29}{48},\frac{19}{24}\right\},\left\{\frac{1}{24},\frac{23}{48},\frac{47}{48}\right\},\left\{\frac{5}{24},\frac{19}{48},\frac{43}{48}\right\}$,
\\
&
$\left\{\frac{11}{48},\frac{13}{24},\frac{35}{48}\right\},\left\{\frac{7}{48},\frac{31}{48},\frac{17}{24}\right\},\left\{\frac{7}{24},\frac{17}{48},\frac{41}{48}\right\},\left\{\frac{13}{48},\frac{11}{24},\frac{37}{48}\right\}$
\end{tabular}
~
\begin{tabular}{c|c}
$n$ & 3d $\ell=3$ exponents
\\ \hline
3 & $\left\{ 0, {1\over 3}, {2\over 3}\right\}$
\\
4 &$ \left\{0,\frac{1}{4},\frac{3}{4}\right\}$
\\
5 & $\left\{ 0, {1\over 5}, {4\over 5}\right\} , \left\{ 0, {2\over 5}, {3\over 5}\right\} $
\\
7 & $ \left\{ {1\over 7}, {2\over 7}, {4\over 7}\right\}, \left\{ {3\over 7}, {5\over 7}, {6\over 7}\right\} $
\\
8&$ \left\{\frac{1}{8},\frac{1}{4},\frac{5}{8}\right\},\left\{\frac{3}{8},\frac{3}{4},\frac{7}{8}\right\}$
\\
12 & $\left\{{1\over 12}, {1\over 3},{7 \over 12} \right\} , \left\{{5\over 12}, {2\over 3},{11 \over 12} \right\}$
\\
15 &$ \left\{{2\over 15}, {1\over 3}, {8 \over 15} \right\} , \left\{{1\over 3}, {11\over 15}, {14 \over 15} \right\} , \left\{{7\over 15}, {2\over 3}, {13 \over 15} \right\} , \left\{{1\over 15}, {4\over 15}, {2 \over 3} \right\} $
\\
16 &$ \left\{{1\over 16}, {3\over 8},{9 \over 16} \right\} ,  \left\{{1\over 8}, {3\over 16},{11 \over 16} \right\},   \left\{{5\over 16}, {13\over 16},{7 \over 8} \right\} ,  \left\{{7\over 16}, {5\over 8},{15 \over 16} \right\} $
\\
21 &$ \left\{{1\over 21}, {4\over 21}, {16\over 21} \right\}  ,  \left\{{2\over 21}, {8\over 21}, {11\over 21} \right\} , \left\{{5\over 21}, {17\over 21}, {20\over 21} \right\} , \left\{{10\over 21}, {13\over 21}, {19\over 21} \right\}$
\\
24 &  $ \left\{{1 \over 24}, {5 \over 12}, {13 \over 24} \right\} , \left\{{1 \over 12}, {5 \over 24}, {17 \over 24} \right\} ,\left\{{7 \over 24}, {19 \over 24}, {11 \over 12} \right\} , \left\{{11 \over 24}, {7 \over 12}, {23 \over 24} \right\} $
\\
48 & $ \left\{{1\over 48}, {11\over 24 }, {25 \over 48} \right\}, \left\{{1\over 24}, {11\over 48 }, {35 \over 48} \right\} , \left\{{19\over 48}, {17\over 24 }, {43 \over 48} \right\} , \left\{{23\over 48}, {13\over 24 }, {47 \over 48} \right\} $,
\\
&
$\left\{{7\over 48}, {5\over 24 }, {31 \over 48} \right\} , \left\{{17\over 48}, {19\over 24 }, {41 \over 48} \right\}  , \left\{{5\over 48}, {7\over 24 }, {29 \over 48} \right\} , \left\{{13\over 48}, {37\over 48 }, {23 \over 24} \right\} $
\end{tabular}
}
}
\end{center}
\caption{Possible exponents mod 1 for three-character theories, disallowing duplications.}
\label{tab:3dl3}
\end{table}

\begin{table}[H]
\begin{center}
\scalebox{.9}{
\makebox[\textwidth]{
\begin{tabular}{c|c}
$n$ & 4d $\ell=0$ exponents
\\ \hline
5&$ \left\{\frac{1}{5},\frac{2}{5},\frac{3}{5},\frac{4}{5}\right\}$
\\
8&$ \left\{\frac{1}{8},\frac{3}{8},\frac{5}{8},\frac{7}{8}\right\}$
\\
9 &$ \left\{\frac{1}{9},\frac{4}{9},\frac{2}{3},\frac{7}{9}\right\},  \left\{\frac{2}{9},\frac{1}{3},\frac{5}{9},\frac{8}{9}\right\} $
\\
10&$  \left\{\frac{1}{10},\frac{3}{10},\frac{7}{10},\frac{9}{10}\right\}$
\\
12 &$ \left\{\frac{1}{12},\frac{5}{12},\frac{7}{12},\frac{11}{12}\right\} $
\\
15 &$ \left\{\frac{2}{15},\frac{7}{15},\frac{8}{15},\frac{13}{15}\right\}, \left\{\frac{1}{15},\frac{4}{15},\frac{11}{15},\frac{14}{15}\right\}$
\\
18 &$ \left\{\frac{1}{6},\frac{5}{18},\frac{11}{18},\frac{17}{18}\right\},  \left\{\frac{1}{18},\frac{7}{18},\frac{13}{18},\frac{5}{6}\right\}$
\\
20 &$ \left\{\frac{1}{20},\frac{9}{20},\frac{13}{20},\frac{17}{20}\right\}, \left\{\frac{3}{20},\frac{7}{20},\frac{11}{20},\frac{19}{20}\right\},\left\{\frac{1}{20},\frac{9}{20},\frac{11}{20},\frac{19}{20}\right\},\left\{\frac{3}{20},\frac{7}{20},\frac{13}{20},\frac{17}{20}\right\} $
\\
24 &$  \left\{\frac{5}{24},\frac{11}{24},\frac{13}{24},\frac{19}{24}\right\} , \left\{\frac{1}{24},\frac{7}{24},\frac{17}{24},\frac{23}{24}\right\} $
\\
28 &$ \left\{\frac{5}{28},\frac{13}{28},\frac{17}{28},\frac{3}{4}\right\}, \left\{\frac{1}{28},\frac{9}{28},\frac{3}{4},\frac{25}{28}\right\}, \left\{\frac{3}{28},\frac{1}{4},\frac{19}{28},\frac{27}{28}\right\}, \left\{\frac{1}{4},\frac{11}{28},\frac{15}{28},\frac{23}{28}\right\}$
\\
30& $\left\{\frac{1}{30},\frac{11}{30},\frac{19}{30},\frac{29}{30}\right\}, \left\{\frac{7}{30},\frac{13}{30},\frac{17}{30},\frac{23}{30}\right\}$
\\
36 &$ \left\{\frac{1}{36},\frac{13}{36},\frac{25}{36},\frac{11}{12}\right\},  \left\{\frac{7}{36},\frac{5}{12},\frac{19}{36},\frac{31}{36}\right\},\left\{\frac{5}{36},\frac{17}{36},\frac{7}{12},\frac{29}{36}\right\},  \left\{\frac{1}{12},\frac{11}{36},\frac{23}{36},\frac{35}{36}\right\}$
\\
40 & \hspace{27pt} $ \left\{\frac{3}{40},\frac{13}{40},\frac{27}{40},\frac{37}{40}\right\},  \left\{\frac{7}{40},\frac{17}{40},\frac{23}{40},\frac{33}{40}\right\}, \left\{\frac{11}{40},\frac{19}{40},\frac{21}{40},\frac{29}{40}\right\}, \left\{\frac{1}{40},\frac{9}{40},\frac{31}{40},\frac{39}{40}\right\}$\hspace{31pt}
\end{tabular}
}
}

\vspace{10pt}

\scalebox{.9}{
\makebox[\textwidth]{
\begin{tabular}{c|c}
$n$ & 4d $\ell=2$ exponents
\\ \hline
9 & $   \left\{0,\frac{2}{9},\frac{5}{9},\frac{8}{9}\right\} , \left\{\frac{1}{9},\frac{1}{3},\frac{4}{9},\frac{7}{9}\right\}$
\\
12 & $ \left\{\frac{1}{12},\frac{1}{4},\frac{7}{12},\frac{3}{4}\right\} $
\\
15 & $ \left\{\frac{1}{15},\frac{4}{15},\frac{7}{15},\frac{13}{15}\right\},  \left\{\frac{2}{15},\frac{1}{5},\frac{8}{15},\frac{4}{5}\right\} ,  \left\{\frac{2}{5},\frac{3}{5},\frac{11}{15},\frac{14}{15}\right\} $
\\
18 & $ \left\{\frac{5}{18},\frac{11}{18},\frac{5}{6},\frac{17}{18}\right\} ,  \left\{\frac{1}{18},\frac{7}{18},\frac{1}{2},\frac{13}{18}\right\}$
\\
24 & $ \left\{\frac{1}{24},\frac{7}{24},\frac{13}{24},\frac{19}{24}\right\} , \left\{\frac{1}{8},\frac{5}{24},\frac{11}{24},\frac{7}{8}\right\} ,  \left\{\frac{3}{8},\frac{5}{8},\frac{17}{24},\frac{23}{24}\right\} $
\\
30 & $ \left\{\frac{11}{30},\frac{17}{30},\frac{23}{30},\frac{29}{30}\right\},  \left\{\frac{1}{30},\frac{3}{10},\frac{19}{30},\frac{7}{10}\right\} ,\left\{\frac{1}{10},\frac{7}{30},\frac{13}{30},\frac{9}{10}\right\}$
\\
36 &$ \left\{\frac{1}{36},\frac{13}{36},\frac{7}{12},\frac{25}{36}\right\} , \left\{\frac{1}{12},\frac{7}{36},\frac{19}{36},\frac{31}{36}\right\},  \left\{\frac{5}{36},\frac{1}{4},\frac{17}{36},\frac{29}{36}\right\},  \left\{\frac{11}{36},\frac{23}{36},\frac{3}{4},\frac{35}{36}\right\} $
\\
60 & $\left\{\frac{7}{60},\frac{19}{60},\frac{31}{60},\frac{43}{60}\right\} , \left\{\frac{1}{60},\frac{13}{60},\frac{37}{60},\frac{49}{60}\right\} ,   \left\{\frac{7}{60},\frac{13}{60},\frac{37}{60},\frac{43}{60}\right\} ,  \left\{\frac{1}{60},\frac{19}{60},\frac{31}{60},\frac{49}{60}\right\}$
\\
84 & $\left\{\frac{11}{84},\frac{23}{84},\frac{5}{12},\frac{71}{84}\right\}, \left\{\frac{5}{12},\frac{47}{84},\frac{59}{84},\frac{83}{84}\right\} ,\left\{\frac{29}{84},\frac{53}{84},\frac{65}{84},\frac{11}{12}\right\}, \left\{\frac{5}{84},\frac{17}{84},\frac{41}{84},\frac{11}{12}\right\}  $
\\
120 & $\left\{\frac{41}{120},\frac{71}{120},\frac{89}{120},\frac{119}{120}\right\} , \left\{\frac{11}{120},\frac{29}{120},\frac{59}{120},\frac{101}{120}\right\}, \left\{\frac{17}{120},\frac{23}{120},\frac{47}{120},\frac{113}{120}\right\},  \left\{\frac{53}{120},\frac{77}{120},\frac{83}{120},\frac{107}{120}\right\} $
\end{tabular}
}
}

\vspace{10pt}

\scalebox{.9}{
\makebox[\textwidth]{
\begin{tabular}{c|c}
$n$ & 4d $\ell=4$ exponents
\\ \hline
9 & $  \left\{0,\frac{1}{9},\frac{4}{9},\frac{7}{9}\right\}, \left\{\frac{2}{9},\frac{5}{9},\frac{2}{3},\frac{8}{9}\right\}$
\\
12 & $\left\{\frac{1}{4},\frac{5}{12},\frac{3}{4},\frac{11}{12}\right\} $
\\
15 & $  \left\{\frac{1}{15},\frac{4}{15},\frac{2}{5},\frac{3}{5}\right\}, \left\{\frac{2}{15},\frac{8}{15},\frac{11}{15},\frac{14}{15}\right\},  \left\{\frac{1}{5},\frac{7}{15},\frac{4}{5},\frac{13}{15}\right\} $
\\
18 & $ \left\{\frac{1}{18},\frac{1}{6},\frac{7}{18},\frac{13}{18}\right\}, \left\{\frac{5}{18},\frac{1}{2},\frac{11}{18},\frac{17}{18}\right\}  $
\\
24 & $\left\{\frac{1}{8},\frac{13}{24},\frac{19}{24},\frac{7}{8}\right\} , \left\{\frac{1}{24},\frac{7}{24},\frac{3}{8},\frac{5}{8}\right\} ,  \left\{\frac{5}{24},\frac{11}{24},\frac{17}{24},\frac{23}{24}\right\} $
\\
30 & $\left\{\frac{1}{30},\frac{7}{30},\frac{13}{30},\frac{19}{30}\right\},  \left\{\frac{3}{10},\frac{11}{30},\frac{7}{10},\frac{29}{30}\right\} , \left\{\frac{1}{10},\frac{17}{30},\frac{23}{30},\frac{9}{10}\right\} $
\\
36 & $ \left\{\frac{5}{36},\frac{17}{36},\frac{29}{36},\frac{11}{12}\right\}, \left\{\frac{11}{36},\frac{5}{12},\frac{23}{36},\frac{35}{36}\right\} , \left\{\frac{1}{36},\frac{1}{4},\frac{13}{36},\frac{25}{36}\right\}, \left\{\frac{7}{36},\frac{19}{36},\frac{3}{4},\frac{31}{36}\right\} $
\\
60 & $ \left\{\frac{11}{60},\frac{23}{60},\frac{47}{60},\frac{59}{60}\right\} ,  \left\{\frac{17}{60},\frac{29}{60},\frac{41}{60},\frac{53}{60}\right\},  \left\{\frac{17}{60},\frac{23}{60},\frac{47}{60},\frac{53}{60}\right\},  \left\{\frac{11}{60},\frac{29}{60},\frac{41}{60},\frac{59}{60}\right\} $
\\
84 & $\left\{\frac{1}{12},\frac{43}{84},\frac{67}{84},\frac{79}{84}\right\} , \left\{\frac{1}{12},\frac{19}{84},\frac{31}{84},\frac{55}{84}\right\},  \left\{\frac{1}{84},\frac{25}{84},\frac{37}{84},\frac{7}{12}\right\} , \left\{\frac{13}{84},\frac{7}{12},\frac{61}{84},\frac{73}{84}\right\}$
\\
120 & $ \left\{\frac{1}{120},\frac{31}{120},\frac{49}{120},\frac{79}{120}\right\} , \left\{\frac{19}{120},\frac{61}{120},\frac{91}{120},\frac{109}{120}\right\} ,  \left\{\frac{7}{120},\frac{73}{120},\frac{97}{120},\frac{103}{120}\right\} ,  \left\{\frac{13}{120},\frac{37}{120},\frac{43}{120},\frac{67}{120}\right\} $
\end{tabular}
}
}
\end{center}
\caption{Possible exponents mod 1 for four-character theories, disallowing duplications.}
\label{tab:4dl2}
\end{table}

\begin{table}[H]
\begin{center}
\scalebox{.9}{
\makebox[\textwidth]{
\begin{tabular}{c|c}
$n$ & 5d $\ell=0$ exponents
\\\hline
15 & $ \left\{\frac{1}{3},\frac{2}{15},\frac{8}{15},\frac{11}{15},\frac{14}{15}\right\} $
\\
33 &$ \left\{\frac{2}{33},\frac{8}{33},\frac{17}{33},\frac{29}{33},\frac{32}{33}\right\},  \left\{\frac{5}{33},\frac{14}{33},\frac{20}{33},\frac{23}{33},\frac{26}{33}\right\}$
\end{tabular}
\begin{tabular}{c|c }
$n$ & 5d $\ell=1$ exponents
\\ \hline
10 &$ \left\{\frac{1}{10},\frac{3}{10},\frac{1}{2},\frac{7}{10},\frac{9}{10}\right\}$
\\
22 &$\left\{\frac{1}{22},\frac{3}{22},\frac{5}{22},\frac{9}{22},\frac{15}{22}\right\} , \left\{\frac{7}{22},\frac{13}{22},\frac{17}{22},\frac{19}{22},\frac{21}{22}\right\} $
\end{tabular}
}
}

\vspace{10pt}

\scalebox{.9}{
\makebox[\textwidth]{
\begin{tabular}{c|c}
$n$ & 5d $\ell=2$ exponents
\\ \hline
15 &$ \left\{\frac{1}{15},\frac{4}{15},\frac{7}{15},\frac{2}{3},\frac{13}{15}\right\} $
\\
33 &$ \left\{\frac{7}{33},\frac{10}{33},\frac{13}{33},\frac{19}{33},\frac{28}{33}\right\} , \left\{\frac{1}{33},\frac{4}{33},\frac{16}{33},\frac{25}{33},\frac{31}{33}\right\} $
\end{tabular}
\begin{tabular}{c|c}
$n$ & 5d $\ell=3$ exponents
\\ \hline
30 &$ \left\{\frac{1}{30},\frac{7}{30},\frac{13}{30},\frac{19}{30},\frac{5}{6}\right\} $
\\
66 &$  \left\{\frac{1}{66},\frac{25}{66},\frac{31}{66},\frac{37}{66},\frac{49}{66}\right\},   \left\{\frac{7}{66},\frac{13}{66},\frac{19}{66},\frac{43}{66},\frac{61}{66}\right\}$
\end{tabular}
}
}

\vspace{10pt}

\scalebox{.9}{
\makebox[\textwidth]{
\begin{tabular}{c|c}
$n$ & 5d $\ell=4$ exponents
\\ \hline
5 &$ \left\{0,\frac{1}{5},\frac{2}{5},\frac{3}{5},\frac{4}{5}\right\}$
\\
11 &$ \left\{\frac{2}{11},\frac{6}{11},\frac{7}{11},\frac{8}{11},\frac{10}{11}\right\}, \left\{\frac{1}{11},\frac{3}{11},\frac{4}{11},\frac{5}{11},\frac{9}{11}\right\}$
\end{tabular}
\begin{tabular}{c|c}
$n$ & 5d $\ell=5$ exponents
\\ \hline
30 &$ \left\{\frac{1}{6},\frac{11}{30},\frac{17}{30},\frac{23}{30},\frac{29}{30}\right\} $
\\
66 &$  \left\{\frac{5}{66},\frac{23}{66},\frac{47}{66},\frac{53}{66},\frac{59}{66}\right\}, \left\{\frac{17}{66},\frac{29}{66},\frac{35}{66},\frac{41}{66},\frac{65}{66}\right\}$
\end{tabular}
}
}
\end{center}
\caption{Possible exponents mod 1 for five-character theories, disallowing duplications.}
\label{tab:5dl1}
\end{table}

\section{Further denominator-rank constraints}
\label{app:furtherdenom}
\subsection{Possible projective denominators for given rank}

Define the projective denominator $N$ to be the least common denominator of the weights $\{h_i\}$ instead of the exponents $\{h_i-\frac{c}{24}\}$.  Like $n$, a denominator-rank constraint for $N$ can be be deduced.  First, by simply ignoring factors of one-dimensional representations, the possible values of $N$ given the rank $d$ must belong to the set
\ie
\mathfrak{Den}'(d) = \bigcup_{f \in \cF} \mathrm{lcm} \left[ \mathfrak{den}_0(\delta_1^{(f)}), \dots,  \mathfrak{den}_0(\delta_{|f|}^{(f)}) \right] \, ,
\fe
where $\mathfrak{den}_0$ and $\cF$ were both defined near \eqref{den0}.
Explicitly, we find
\ie
\mathfrak{Den}'(1) &= \{ 1 \} \, ,
\\
\mathfrak{Den}'(2) &= \{ 2, 3, 4, 5, 8 \} \, ,
\\
\mathfrak{Den}'(3) &= \{ 3, 4, 5, 7, 8, 16 \} \, ,
\\
\mathfrak{Den}'(4) &= \{ 2, 3, 4, 5, 6, 7, 8, 9, 10, 12, 15, 20, 24, 40 \} \, ,
\\
\mathfrak{Den}'(5) &= \{ 5, 11 \} \, .
\fe
However, some of the above denominators can be further eliminated.  If every order $N$ element acts trivially on every $d$-dimensional irreducible projective representation of $\PSL(2,\Z_N)$, then the actual projective denominator is a proper factor of $N$.  In practice, we consider the Schur cover of $\PSL(2,\Z_N)$, look at every $d$-dimensional linear representation, and compute the projective kernel (the set of conjugacy classes whose characters are related to the trivial character by roots of unity).  We then quotient by the projective kernel, and examine whether the quotient group has any order $N$ element.  If none of the $d$-dimensional representations gives a quotient group that has an order $N$ element, then this $N$ is not possible for this particular $d$.  This eliminates $N=8$ from $d=2$, and $N=2,3,8$ from $d=4$.  We find\footnote{Note that $N = 24, 40$ for $\mathfrak{Den}(4)$ may be potentially ruled out, but the GAP \cite{GAP4} computations did not complete in a reasonable amount of time. We thus leave them in as possible denominators.}
\ie
\mathfrak{Den}(1) &= \{ 1 \} \, ,
\\
\mathfrak{Den}(2) &= \{ 2, 3, 4, 5 \} \, ,
\\
\mathfrak{Den}(3) &= \{ 3, 4, 5, 7, 8, 16 \} \, ,
\\
\mathfrak{Den}(4) &= \{ 4, 5, 6, 7, 9, 10, 12, 15, 20, 24, 40 \} \, ,
\\
\mathfrak{Den}(5) &= \{ 5, 11 \} \, .
\fe

\subsection{Lower bound on rank for given denominator}

It is clear that a lower bound on the rank realizing a certain denominator $n$ (of exponents $\{h_i-\frac{c}{24}\}$) can be inferred from Table~\ref{Tab:SL2ZNRep}.  For instance, a lower bound on the rank for $n = 11$ is given by $\frac12(11-1) = 5$, and a lower bound on the rank for $n = 13$ is given by $\frac12(13+1) = 7$.\footnote{For $n = 13$, notice that $\frac12(13-1) = 6$ is not allowed because $\sigma = -$.
}
It is clear that the lower bound is trivial
for any $n \in \{ 1, 2, 3, 6 \}$.

If we are given $N$ instead of $n$, we can also derive a lower bound.
Suppose $N$ is odd, then a Schur cover of $\PSL(2,\Z_N)$ is given by $\SL(2,\Z_N)$.  Let $N = \prod_{i=1}^{n_p} p_i^{\lambda_i}$ be the prime factor decomposition of $N$, which involves $n_p$ distinct primes.  The linear representation $\rho : \SL(2,\Z_N) \to \GL(d,\CC)$ is the tensor product of linear representations $\bigotimes_i \rho_i$, where $\rho_i : \SL(2,p_i^{\lambda_i}) \to \GL(d,\CC)$ is such that $\ker p \circ \rho_i$ does not contain all the order $p_i^{\lambda_i}$ conjugacy classes (equivalently, $\SL(2,p_i^{\lambda_i})/\ker p \circ \rho_i$ has an order $N$ element).  Clearly, each $\rho_i$ cannot be one-dimensional, so we automatically obtain a lower bound of $\text{rank} \ge 2^{n_p}$.  Taking Table~\ref{Tab:SL2ZNRep} into account, we find (the max is just to take care of the special case of $p = 3, \lambda = 1$)
\ie
\text{rank} \ge \prod_{i=1}^{n_p} \, \text{max} \left( \frac{p_i-1}{2} , 2 \right) \times \lfloor (p_i+1) p_i^{\lambda_i - 2} \rfloor \, .
\fe
For instance, if $N = 3 \times 5 \times 7 = 105$, then
\ie
\text{rank} \ge 2 \times 2 \times 3 = 12 \, .
\fe

\section{Finite subgroups of $\SL(3,\CC)$ and $\SL(4,\CC)$}
\label{app:SLdfinsubgs}
In this appendix we summarize the finite subgroups of  $\SL(3,\CC)$ and $\SL(4,\CC)$. In both cases, the finite subgroups can be split into three classes: the intransitive subgroups (analogs of cyclics groups of $\SL(2,\CC)$), the transitive imprimitive subgroups (analogs of the dihedral subgroups of $\SL(2,\CC)$, and the transitive primitive subgroups (analogs of the exceptional finite subgroups of $\SL(2,\CC)$). Here we discuss only the transitive subgroups.

\subsection{Finite subgroups of $\SL(3,\CC)$}
\label{app:SL3subs}
We begin with the finite subgroups of $\SL(3,\CC)$. We consider the cases of transitive primitive and transitive imprimitive subgroups separately.

\subsubsection*{Transitive primitive subgroups}
There are a total of eight transitive primitive subgroups, given as follows. First, defining
\bea
S&=& \left( \begin{matrix}  1 &0 &0 \\ 0 & \omega & 0 \\ 0 &0 & \omega^2 \end{matrix}\right)~, \hspace{0.5 in} T =  \left(\begin{matrix} 0 & 1 & 0 \\ 0 &0 & 1 \\ 1 &0 & 0  \end{matrix} \right)~,
\no\\
U &=& \left(\begin{matrix}\xi_9^2 & 0 & 0 \\ 0 & \xi_9^2 & 0 \\ 0 & 0 & \xi_9^2\, \omega \end{matrix} \right)~, \hspace{0.5 in}  V =  {1 \over i \sqrt{3}} \left(\begin{matrix} 1& 1 & 1 \\ 1 & \omega & \omega^2 \\ 1 & \omega^2 & \omega \end{matrix} \right)~,
\eea
where $\xi_n := e^{2 \pi i /n}$ and $\omega := \xi_3$. Then we have the following three groups
\bea
G_I^{(3)} &:=& \langle S, \, T, \, V, U \rangle~, \hspace{0.9 in} |G_I^{(3)}| = 648
\no\\
G_{II}^{(3)} &:=& \langle S, \, T, \, V, U V U^{-1} \rangle~,\hspace{0.5 in} |G_{II}^{(3)}| = 216
\no\\
G_{III}^{(3)} &:=& \langle S, \, T, \, V \rangle~,\hspace{1.1 in} |G_{III}^{(3)}| = 108~.
\eea
The group $G_{II}^{(3)}$ is known as the \textit{Hesse group}, and the group $G_{I}^{(3)}$ is a triple cover of it.

Next there is a group of order 60,
\bea
G_{IV}^{(3)} &:=& \left\langle \left(\begin{matrix} 1 &0 & 0 \\ 0 & \xi_5^4 & 0 \\ 0 &0 & \xi_5 \end{matrix} \right), \, \left(\begin{matrix} -1 &0 & 0 \\ 0 &0 & -1\\ 0 & -1 &0 \end{matrix} \right)  , \,{1 \over \sqrt{5}} \left(\begin{matrix} 1 & 1 & 1 \\ 2 & s & t\\ 2 & t & s \end{matrix} \right)  \right\rangle~,\hspace{0.5 in} |G_{IV}^{(3)}| = 60
\eea
where $s = \xi_5^2 + \xi_5^3$ and $t = \xi_5 + \xi_5^4$. This group is isomorphic to the icosahedral group $A_5$. There is also a trivial $\ZZ_3$ extension of this group,
\bea
G_{V}^{(3)} &:=& G_{IV}^{(3)}  \times \ZZ_3~,\hspace{0.8 in} |G_{V}^{(3)}| = 180~.
\eea

Next one has the group
\bea
G_{VI}^{(3)} : = \left\langle \left( \begin{matrix} \xi_7 & 0 & 0 \\ 0 & \xi_7^2 &0 \\ 0 & 0 & \xi_7^4\end{matrix}\right),\, \left(\begin{matrix} 0 & 1 & 0 \\ 0 &0 & 1 \\ 1 &0 & 0  \end{matrix} \right), \, {i \over \sqrt{7}}\left(\begin{matrix} r & u & v \\ u & v & r \\ v & r & u\end{matrix} \right)  \right \rangle~,\hspace{0.5 in} |G_{VI}^{(3)}| = 168
\eea
where $r = \xi_7^4 - \xi_7^3$, $ u = \xi_7^2 - \xi_7^5$, and $v = \xi_7 - \xi_7^6$. This groups is known as the \textit{Klein group}, and is isomorphic to $G_{VI}^{(3)}\cong \PSL(2,7) \cong \GL(3,2)$. There is also a trivial $\ZZ_3$ extension of this group,
\bea
G_{VII}^{(3)} &:=& G_{VI}^{(3)}  \times \ZZ_3~,\hspace{0.8 in} |G_{VII}^{(3)}| = 504~.
\eea
Finally, there is an extension of $G_{IV}^{(3)}$ by an additional non-trivial generator,
\bea
G_{VIII}^{(3)} &: =& \left\langle G_{IV}^{(3)}\,,\,\, {1 \over \sqrt{5}}\left(\begin{matrix} 1 & \lambda & \lambda \\ 2 \overline \lambda & s & t \\ 2 \overline \lambda & t & s \end{matrix} \right) \right \rangle~,\hspace{0.5 in} |G_{VIII}^{(3)}| = 1080
\eea
where $\lambda : = {1\over 4}(-1 + i \sqrt{15})$. This group is known as the \textit{Valentiner group}, and is a perfect triple cover of $A_6$.

\subsubsection*{Transitive imprimitive subgroups}

We now turn to transitive imprimitive finite subgroups. These are analogous to the dihedral series for $\SL(2,\CC)$, and give two infinite families of monodromy groups.
First, there are groups of the type
\bea
G^{(3)}(a,b) &:=& \left\langle \left(\begin{matrix} a & 0 & 0 \\ 0 & b & 0 \\ 0 & 0 & {1 \over a b}\end{matrix} \right),\,  \left(\begin{matrix} 0 & 1 & 0 \\ 0 &0 & 1 \\ 1 &0 & 0  \end{matrix} \right)
 \right\rangle~.
\eea
for arbitrary $a, b \in \CC$. Second, there are groups obtained by adding an additional generator,
\bea
G^{(3)}(a,b, a', b') &:=& \left\langle \left(\begin{matrix} a & 0 & 0 \\ 0 & b & 0 \\ 0 & 0 & {1 \over a b}\end{matrix} \right),\,  \left(\begin{matrix} 0 & 1 & 0 \\ 0 &0 & 1 \\ 1 &0 & 0  \end{matrix} \right),\,\left(\begin{matrix} 0 & a'& 0 \\ b' & 0 & 0 \\ 0 & 0 & -{1 \over a' b'}\end{matrix} \right)
 \right\rangle~.
\eea
for arbitrary $a, b,a',b' \in \CC$.

\subsection{Finite subgroups of $\SL(4,\CC)$}
\label{app:SL4subs}

The finite subgroups of $\SL(4,\CC)$ have been catalogued in \cite{Hanany:1998sd}. Here we summarize those results, and correct several typographic errors as well. As before, the classification can be divided into two parts: transitive primitive and transitive imprimitive. The former are the analogs of the exceptional series of $\SL(2,\CC)$; there are 30 such cases. The latter are analogs of the dihedral series of $\SL(2,\CC)$; here there are 5 infinite families to be considered.

\subsubsection*{Transitive primitive subgroups}
We begin by presenting the following matrix generators:
\bea
F_1 &=& \begin{pmatrix} 1 & 0 & 0 &0 \\ 0 & 1 & 0 & 0 \\ 0 & 0 & \omega & 0 \\ 0 & 0 & 0 & \omega^2\end{pmatrix}\hspace{0.5 in} F_2 ={1\over \sqrt{3}} \begin{pmatrix}1 & 0 & 0 & \sqrt{2} \\ 0 & -1 & \sqrt{2} & 0 \\ 0& \sqrt{2} & 1 & 0 \\ \sqrt{2} & 0 & 0 &-1 \end{pmatrix} \hspace{0.5 in} F_3 = \begin{pmatrix}{\sqrt{3} \over 2} & \half & 0 & 0 \\ \half & - {\sqrt{3}\over 2}& 0 & 0 \\ 0 & 0 & 0 & 1 \\ 0 & 0 & 1 & 0 \end{pmatrix}
\no\\
F_1 &=&{1\over 3} \begin{pmatrix} 3& 0 & 0 & 0 \\ 0 & -1 & 2 & 2 \\ 0 & 2 & -1& 2 \\ 0 & 2 & 2 & -1 \end{pmatrix}\hspace{0.5 in}F_3'. = {1\over 4}\begin{pmatrix} -1 & \sqrt{15} & 0 & 0 \\ \sqrt{15} & 1 & 0 & 0 \\ 0 & 0 & 0 & 4\\ 0 & 0 & 4 & 0\end{pmatrix}\hspace{0.5 in}F_4 = \begin{pmatrix}0 & 1 & 0 & 0 \\ 1 & 0 & 0 & 0 \\ 0 & 0 & 0 & -1 \\ 0 & 0 & -1 & 0 \end{pmatrix}
\no\\
S & = & \begin{pmatrix}1 & 0 & 0 & 0 \\ 0 & \xi_7 & 0 & 0 \\ 0 & 0 & \xi_7^4 & 0 \\ 0 & 0 & 0 & \xi_7^2  \end{pmatrix} \hspace{0.5 in} T = \begin{pmatrix} 1 & 0 & 0 & 0 \\ 0 & 0 & 1 & 0 \\ 0 & 0 & 0 & 1 \\ 0 & 1 & 0 & 0 \end{pmatrix} \hspace{0.5 in} W = {1\over i \sqrt{7}}\begin{pmatrix} p^2 & 1 & 1 & 1 \\ 1 & -q & -p& -p \\ 1& -p&-q& -p\\ 1 & -p&-p&-q\end{pmatrix}
\no\\
R&=& {1\over \sqrt{7}}\begin{pmatrix} 1 & 1 & 1 & 1 \\ 2 & s & t& u \\ 2 & t & u & s \\ 2 & u & s & t\end{pmatrix}\hspace{0.5 in} D = \begin{pmatrix}\omega & 0 & 0 & 0 \\ 0 & \omega & 0 & 0 \\ 0 & 0 & \omega & 0 \\ 0 & 0 & 0 & 1 \end{pmatrix}\hspace{0.5 in} V = {1 \over i \sqrt{3}}\begin{pmatrix} i \sqrt{3} & 0 & 0 & 0 \\ 0 & 1 & 1 & 1 \\ 0 & 1 & \omega & \omega^2 \\ 0 & 1 & \omega^2 & \omega \end{pmatrix}
\no\\
F&=& \begin{pmatrix}0 & 0 & -1 & 0 \\ 0 & 1 & 0 & 0 \\ -1 & 0 & 0 & 0 \\ 0 & 0 & 0 & -1 \end{pmatrix} \hspace{0.5 in} F' = {1 + i \over \sqrt{2}} \begin{pmatrix} 1 & 0 & 0 & 0 \\ 0 & 1 & 0 & 0 \\ 0 & 0 & 0 & 1\\ 0 & 0 & 1 & 0\end{pmatrix} \hspace{0.5in} F'' = \begin{pmatrix} 0 & 1 & 0 & 0 \\ -1 & 0 & 0 & 0 \\ 0 & 0 & 0 & 1 \\ 0 & 0 & -1 & 0 \end{pmatrix}\no
\eea
where $\xi_n := e^{2 \pi i /n}$ and $\omega = \xi_3$, $p = \xi_7 + \xi_7^2 + \xi_7^4$, $q  = \xi_7^3 + \xi_7^5 + \xi_7^6$, $ s= \xi_7^2 + \xi_7^5$, $t = \xi_7^3 + \xi_7^4$, and $u = \xi_7 + \xi_7^6$. In terms of these we may define the following groups,
\bea
G_I^{(4)} &  = & \langle F_1, F_2, F_3 \rangle~, \hspace{1.1 in} |G_I^{(4)}| = 240
\no\\
G_{II}^{(4)} &  = & \langle F_1, F_2', F_3'\rangle~, \hspace{1.1 in} |G_{II}^{(4)}| = 60
\no\\
G_{III}^{(4)} &  = & \langle F_1, F_2, F_3, F_4\rangle~, \hspace{0.85 in} |G_{III}^{(4)}| = 1440
\no\\
G_{IV}^{(4)} &  = & \langle S, T, W \rangle~,  \hspace{1.2 in} |G_{IV}^{(4)}| = 5040
\no\\
G_{V}^{(4)} &=& \langle S, T, R \rangle~,  \hspace{1.25 in} |G_{V}^{(4)}| = 672
\no\\
G_{VI}^{(4)} &  = & \langle F_1, T,  D, V, F \rangle~, \hspace{0.8 in} |G_{VI}^{(4)}| = 51840
\no\\
G_{VII}^{(4)} &  = & \langle G_I^{(4)}, F'' \rangle~, \hspace{1.2 in} |G_{VII}^{(4)}| = 480
\no\\
G_{VIII}^{(4)} &  = & \langle G_{II}^{(4)}, F' \rangle~, \hspace{1.2 in} |G_{VIII}^{(4)}| = 480
\no\\
G_{IX}^{(4)} &  = & \langle G_{III}^{(4)}, F'' \rangle~, \hspace{1.15 in} |G_{IX}^{(4)}| = 2880
\eea
Let us remark on some of these. The group $G_I^{(4)}$ is isomorphic to a double cover of $\SL(2,5)$. The group $G_{II}^{(4)}$ is a four-dimensional representation of $A_5$. The group $G_{III}^{(4)}$ is an extension of $A_6$ by the Abelian center of $SU(4)$, while the group $G_{IV}^{(4)}$ is an extension of $A_7$ by a $\ZZ_2$ subgroup of the center. The group $G_{V}^{(4)}$ is the four-dimensional analog of the Klein group. Finally, the groups $G_{VII}^{(4)},G_{VIII}^{(4)},$ and $G_{IX}^{(4)}$ are central extensions of $S_5$, $S_5$, and $S_6$ respectively.

We next introduce the following $2\times 2$ matrices,
\bea
&\vphantom{.}&S_{SU(2)} = \half \begin{pmatrix} -1+i & -1+ i \\ 1+i & -1-i\end{pmatrix} \hspace{0.5 in} U_{SU(2)} = {1\over \sqrt{2}}\begin{pmatrix}1+i & 0 \\ 0 & 1-i \end{pmatrix}
\no\\
&\vphantom{.}& \hspace{0.5 in}V_{SU(2)} = \begin{pmatrix} {i\over 2} & {1-\sqrt{5} \over 4} - i {1+ \sqrt{5} \over 4}\\ -{1-\sqrt{5} \over 4} - i {1+ \sqrt{5} \over 4} & - {i \over 2}\end{pmatrix} \hspace{0.5 in}
\no\\
 x_1& =& {1\over \sqrt{2}}\begin{pmatrix} 1 & 1 \\ i & - i\end{pmatrix}\hspace{0.5in} x_2 = {1\over \sqrt{2}} \begin{pmatrix} i & i \\ -1 & 1 \end{pmatrix}\hspace{0.5in} x_3 = {1\over \sqrt{2}} \begin{pmatrix}-1 & -1 \\ -1 & 1 \end{pmatrix}
\no\\
x_4 &=& {1\over \sqrt{2}}\begin{pmatrix} i & 1 \\ 1 & i\end{pmatrix} \hspace{0.5 in} x_5 = {1\over \sqrt{2}} \begin{pmatrix} 1 & -1 \\ -i & -i\end{pmatrix}\hspace{0.5 in} x_6 = {1\over \sqrt{2}} \begin{pmatrix} i & -i \\ 1 & 1\end{pmatrix}\no
\eea
in terms of which we may define the following finite subgroups
\bea
G_{X}^{(4)} &=& \langle S_{SU(2)}, U^2_{SU(2)}\rangle \otimes \langle S_{SU(2)}, U^2_{SU(2)}\rangle~,  \hspace{1.2 in} |G_{X}^{(4)} | = 288
\no\\
G_{XI}^{(4)} &=&\langle x_1 \otimes x_2, x_1 \otimes x_2^T, x_3 \otimes x_4, x_5 \otimes x_6\rangle~,\hspace{1.1 in} |G_{XI}^{(4)} | = 576
\no\\
G_{XII}^{(4)} &=&  \langle S_{SU(2)}, U^2_{SU(2)}\rangle \otimes \langle S_{SU(2)}, U_{SU(2)}\rangle~,  \hspace{1.2 in} |G_{XII}^{(4)} | = 576
\no\\
G_{XIII}^{(4)} &=&  \langle S_{SU(2)}, U^2_{SU(2)}\rangle \otimes \langle S_{SU(2)}, V_{SU(2)},U^2_{SU(2)}\rangle~,  \hspace{0.7 in} |G_{XIII}^{(4)} | = 1440
\no\\
G_{XIV}^{(4)} &=&  \langle S_{SU(2)}, U_{SU(2)}\rangle \otimes  \langle S_{SU(2)}, U_{SU(2)}\rangle~,  \hspace{1.2 in} |G_{XIV}^{(4)} | = 1152
\no\\
G_{XV}^{(4)} &=&  \langle S_{SU(2)}, U_{SU(2)}\rangle \otimes  \langle S_{SU(2)}, V_{SU(2)},U^2_{SU(2)}\rangle~,  \hspace{0.7 in} |G_{XV}^{(4)} | = 2880
\no\\
G_{XVI}^{(4)} &=& \langle S_{SU(2)}, V_{SU(2)},U^2_{SU(2)}\rangle \otimes  \langle S_{SU(2)}, V_{SU(2)},U^2_{SU(2)}\rangle~,  \hspace{0.2 in} |G_{XVI}^{(4)} | = 7200
\eea
Note that we have the normal sequence $G_{X}^{(4)}<G_{XI}^{(4)}<G_{XIV}^{(4)}$.
We may combine some of the above groups with the following generators
\bea
T_1 = {1 + i \over \sqrt{2}} \begin{pmatrix} 1& 0 & 0 & 0 \\ 0 & 0 & 1 & 0 \\ 0 & 1 & 0 & 0 \\ 0 & 0 & 0 & 1 \end{pmatrix}\hspace{0.5 in}T_2 = \begin{pmatrix}1& 0 & 0 & 0 \\ 0 & 0 & 1 & 0 \\ 0 & i & 0 & 0 \\ 0 & 0 & 0 & i \end{pmatrix}
\eea
to form the following five additional subgroups,
\bea
G_{XVII}^{(4)} &=& \langle G_{XI}^{(4)}, T_1\rangle~,  \hspace{0.6 in} |G_{XVII}^{(4)} | = 2304
\no\\
G_{XVIII}^{(4)} &=& \langle G_{XI}^{(4)}, T_2\rangle~,  \hspace{0.6 in} |G_{XVIII}^{(4)} | = 2304
\no\\
G_{XIX}^{(4)} &=& \langle G_{X}^{(4)}, T_1\rangle~,  \hspace{0.6 in} |G_{XIX}^{(4)} | = 1152
\no\\
G_{XX}^{(4)} &=& \langle G_{XVI}^{(4)}, T_1\rangle~,  \hspace{0.5 in} |G_{XX}^{(4)} | = 28800
\no\\
G_{XXI}^{(4)} &=& \langle G_{XIV}^{(4)}, T_1\rangle~,  \hspace{0.5 in} |G_{XXI}^{(4)} | = 4608
\eea
Finally, defining the following generators,
\bea
&\vphantom{.}& \hspace{0.8 in}A = {1 + i \over \sqrt{2}} \begin{pmatrix} 1 & 0 & 0 & 0 \\ 0 & i & 0 & 0 \\ 0 & 0 & i & 0 \\ 0 & 0 & 0 & 1\end{pmatrix}  \hspace{0.5 in} B = {1+ i \over \sqrt{2}} \begin{pmatrix} 1 & 0 & 0 & 0 \\ 0 & 1 & 0 & 0 \\ 0 & 0 & 1 & 0 \\ 0 & 0 & 0 & -1\end{pmatrix}
\no\\
S' &=& {1 + i \over \sqrt{2}} \begin{pmatrix}i & 0 & 0 & 0 \\ 0 & i & 0 & 0 \\ 0 & 0 & 1 & 0 \\ 0 & 0 & 0 & 1 \end{pmatrix}\hspace{0.5 in}T'={1+ i \over 2}\begin{pmatrix}-i & 0 & 0 & i \\ 0 & 1 & 1 & 0 \\ 1 & 0 & 0 & 1 \\ 0 & -i & i & 0 \end{pmatrix} \hspace{0.5 in} R' = {1\over \sqrt{2}} \begin{pmatrix} 1 & i & 0 & 0 \\ i & 1 & 0 & 0 \\ 0 & 0 & i & 1 \\ 0 & 0 & -1 & -i\end{pmatrix}\no
\eea
as well as the following group of order $32$,
\bea
K = \left \langle \begin{pmatrix}1 & 0 & 0 & 0 \\ 0 & 1 & 0 & 0 \\ 0 & 0 & -1 & 0 \\ 0 & 0 & 0 & -1 \end{pmatrix}, \begin{pmatrix}1 & 0 & 0 & 0 \\ 0 & -1 & 0 & 0 \\ 0 & 0 & -1 & 0 \\ 0 & 0 & 0 & 1 \end{pmatrix}, \begin{pmatrix} 0 & 1 & 0 & 0 \\ 1 & 0 & 0 & 0 \\ 0 & 0 & 0 & 1 \\ 0 & 0 & 1 & 0\end{pmatrix}, \begin{pmatrix} 0 & 0 & 1 & 0 \\ 0 & 0 & 0 & 1 \\ 1 & 0 & 0 & 0 \\ 0 & 1 & 0 & 0\end{pmatrix} \right \rangle\no
\eea
we may define the last set of primitive subgroups,
\bea
G_{XXII}^{(4)} & = & \langle K, T' \rangle~, \hspace{0.92 in}|G_{XXII}^{(4)}| = 320
\no\\
G_{XXIII}^{(4)} & = & \langle K, T',(R')^2 \rangle~, \hspace{0.5 in}|G_{XXIII}^{(4)}| = 640
\no\\
G_{XXIV}^{(4)} & = & \langle K, T',R' \rangle~, \hspace{0.7 in}|G_{XXIV}^{(4)}| = 1280
\no\\
G_{XXV}^{(4)} & = & \langle K, T',S'B \rangle~, \hspace{0.55 in}|G_{XXV}^{(4)}| = 3840
\no\\
G_{XXVI}^{(4)} & = & \langle K, T',BR' \rangle~, \hspace{0.5 in}|G_{XXVI}^{(4)}| = 3840
\no\\
G_{XXVII}^{(4)} & = & \langle K, T',A \rangle~, \hspace{0.68 in}|G_{XXVII}^{(4)}| = 7680
\no\\
G_{XXVIII}^{(4)} & = & \langle K, T',B \rangle~, \hspace{0.68 in}|G_{XXVIII}^{(4)}| = 7680
\no\\
G_{XXIX}^{(4)} & = & \langle K, T',AB \rangle~, \hspace{0.55 in}|G_{XXIX}^{(4)}| = 23040
\no\\
G_{XXX}^{(4)}& = & \langle K, T',S' \rangle~, \hspace{0.65 in}|G_{XXX}^{(4)}| = 46080
\eea

\subsubsection*{Transitive imprimitive subgroups}

We next present the five infinite families of imprimitive subgroups. The first four can be constructed using the Abelian group
\bea
\label{eq:DeltandefSL4}
\Delta_n = \left\langle \begin{pmatrix}\xi_n^i & 0 & 0 & 0 \\ 0 & \xi_n^j & 0 & 0 \\ 0 & 0 & \xi_n^k & 0 \\ 0 & 0 & 0 & \xi_n^{-(i+j+k)} \end{pmatrix} \big |\,\,\, i,j,k = 1, \dots, n \right \rangle \hspace{0.5 in}  n \in \NN
\eea
of order $n^3$, as well as the four groups $A_4$, $S_4$, $Q_8$ (the dihedral group of 8 elements), and $\ZZ_2 \times \ZZ_2$. We have
\bea
G_1^{(4)}(n) &=& \left\langle \Delta_n, A_4  \right\rangle~,\hspace{0.91 in}|G_1^{(4)}(n)| = 12 n^3
\no\\
G_2^{(4)}(n) &=& \left\langle \Delta_n, S_4  \right\rangle~,\hspace{0.95 in}|G_2^{(4)}(n)| = 24 n^3
\no\\
G_3^{(4)}(n) &=& \left\langle \Delta_n, Q_8 \right\rangle~,\hspace{0.9 in}|G_3^{(4)}(n)| = 8 n^3
\no\\
G_4^{(4)}(n) &=& \left\langle \Delta_n,\ZZ_2 \times \ZZ_2 \right\rangle~,\hspace{0.55 in}|G_4^{(4)}(n)| = 4 n^3
\eea
Finally, there is a remaining infinite family of the form
\bea
G^{(4)}(n; a,b,c,d,e,f) =\left \langle \begin{pmatrix} a & b & 0 & 0 \\ c & {1+b c \over a}& 0 & 0 \\ 0 & 0 & d & e \\ 0 & 0 & f &  {1+ef\over d} \end{pmatrix} , \begin{pmatrix}0 & 0 & 1 & 0 \\ 0 & 0 & 0 & 1 \\ \xi_n^i & 0 & 0 & 0 \\ 0 & \xi_n^{-i} & 0 & 0  \end{pmatrix} \,\,\big | \,\,\, i=1, \dots n\right\rangle
\eea
where $n \in \NN$ and $a,b,c,d,e,f\in \CC$.

\section{Exponents for Imprimitive subgroups of $\SL(4,\CC)$}
\label{app:4dimprim}

In this appendix we derive the set of allowed exponents (\ref{eq:4dimprimres}) quoted for imprimitive subgroups of type $G_i^{(4)}(n)$ for $i=1, \dots, 4$. Since all of these groups are subgroups of $G_2^{(4)}(n)$, we can restrict ourselves to that case. Recall that  $G_2^{(4)}(n)$ is given by $G_2^{(4)}(n) = \langle \Delta_n, S_4 \rangle$, with $\Delta_n$ as defined in (\ref{eq:DeltandefSL4}).

The group $S_4$ has 24 elements, with 5 conjugacy classes. Two of these classes contain 2-cycles, while one class contains 3-cycles. Of the two classes of 2-cycles, only one has trace 0 and is fit for identifying with $S$---namely, the conjugacy class consisting of $\{(12)(34), (13)(24), (14)(23)\}$). On the other hand, all of the 3-cycles (there are 8 of them) have trace 1 and can be identified with $ST$. If we take both $S$ and $ST$ to be elements of $S_4$, there are then 24 possible choices, and hence 24 possible choices for $T$. For example, we can identify $S$ with $(12)(34)$ and $ST$ with $(123)$, which as $4\times 4$ matrices looks like
\bea
S = \begin{pmatrix} 0 & 1 & 0 & 0 \\ 1 & 0 & 0 & 0 \\ 0 & 0 & 0 & 1 \\ 0 & 0 & 1 & 0  \end{pmatrix} \hspace{0.5 in} ST = \begin{pmatrix} 0 & 0 & 1 & 0 \\ 1 & 0 & 0 & 0 \\ 0 & 1 & 0 & 0 \\ 0 & 0 & 0 & 1\end{pmatrix}
\eea
From this we conclude that
\bea
T = \begin{pmatrix} 1 & 0 & 0 & 0 \\ 0 & 0 & 1 & 0 \\ 0 & 0 & 0 & 1\\ 0 & 1 & 0 & 0 \end{pmatrix}
\eea
which has eigenvalues $\{1, 1, e^{2 \pi i \over 3},e^{4 \pi i \over 3}\}$, and thus we read off the possible exponents $\{0,0,{1\over 3}, {2\over 3}\}\,\,{\mathrm{mod} \,\,1}$. In fact, it is easy to show that all 24 choices give rise to this same set of exponents.

Next we consider $S$ or $ST$ taking values in $\Delta_n$, or alternatively being a product of $\Delta_n$ and an element of $S_4$. There are then 7 remaining cases,
\begin{enumerate}
\item $S=[S_4], ST = [\Delta]$
\item $S=[S_4], ST=[\Delta.S_4]$
\item $S=[\Delta], ST=[S_4]$
\item $S=[\Delta], ST=[\Delta.S_4]$
\item $S=[\Delta.S_4], ST = [S_4]$
\item $S=[\Delta.S_4], ST = [\Delta]$
\item $S=[\Delta.S_4], ST = [\Delta.S_4]$
\end{enumerate}
Note that we have not included $S=ST=[\Delta]$, since in that case both $S$ and $ST$ are diagonal, and thus the resulting group is intransitive.

One must now analyze each of the above seven cases. In case 1, $S$ again takes one of the 3 values in $S_4$, but $ST$ is now identified with an element of $\Delta_n$ labelled by $(n,i,j,k)$. Demanding that $(ST)^3=1$ and $\mathrm{Tr}\, ST=1$ restricts $n$ to be a multiple of 3, and restricts $(i,j,k)$ to be one of three combinations:
\bea
(i,j,k) = \left(n,{n\over 3},{2n\over3}\right), \left(n,n,{n\over 3}\right), \left(n,n,{2n\over 3}\right)
\eea
For a given $n$, there are then three possible choices for $ST$. Thus for any $n\in 3\ZZ$ there are a total of $3 \times 3=9$ possible choices for $T$, and one finds by explicit computation that the possible exponents are always either $\left\{0,0, \half, \half \right\}$ or $\left\{{1\over 3},{2\over 3}, {1\over 6}, {5 \over 6}\right \}$. These either have duplications or have been already obtained in the primitive case, so we do not consider them further here.

Moving on to the second case, $ST$ is taken to be a product of an element of $\Delta_n$ and one of the 8 3-cycles. Depending on which of the 8 3-cycles we choose, imposing $(ST)^3=1$ and $\mathrm{Tr}\, ST=1 $ gives different conditions on $i,j,k$ (though in all cases $n$ is unconstrained). When the dust settles, one finds 12 possible classes of $ST$, each of which contain $n^2$ elements labelled by $(p,q)$ such that $p,q=1,\dots,n$.
Computing all $ 3 \times 12 = 36$ possible $T$ matrices, we find that the resulting exponents are of the form
\bea
\left\{ {p \over n},\,\, -{p\over3n}, \,\,{2\over 3} - {p \over 3n},\,\, {1\over 3} - {p\over 3n} \right\}
\eea
where $p=1,\dots,n$ or $p= -2n,\dots, -2$. In this way one proceeds for the remaining five cases. The final result is that the possible new exponents are all of the above form,
where now $p$ can take values anywhere in the range $[-2n, 2n]$.

\bibliography{refs}
\bibliographystyle{JHEP}

\end{document}